\documentclass[referee]{raa}            

\usepackage{graphicx,times}             
\usepackage{natbib}
\usepackage{amssymb,amsmath}
\graphicspath{{./}{figures/}}
\bibpunct{(}{)}{;}{a}{}{,}
\usepackage{listings}
\definecolor{codegreen}{rgb}{0,0.6,0}
\definecolor{codegray}{rgb}{0.5,0.5,0.5}
\definecolor{codepurple}{rgb}{0.58,0,0.82}
\definecolor{backcolour}{rgb}{0.95,0.95,0.92}
\lstdefinestyle{mystyle}{
    backgroundcolor=\color{backcolour},   
    commentstyle=\color{codegreen},
    keywordstyle=\color{black},
    numberstyle=\tiny\color{black},
    stringstyle=\color{codegreen},
    basicstyle=\ttfamily\footnotesize,
    breakatwhitespace=false,         
    breaklines=true,                 
    captionpos=b,                    
    keepspaces=true,                                   
    numbersep=5pt,                  
    showspaces=false,                
    showstringspaces=false,
    showtabs=false,                  
    tabsize=2,
    language=Python
}
\usepackage[pagebackref=true]{hyperref}
\usepackage{xcolor}
\usepackage{multirow}
\usepackage{lscape}
\usepackage{rotating}
\usepackage{multicol}        
\usepackage{bm}		
\usepackage{pdflscape}	
\usepackage{longtable}
\usepackage{threeparttable}
\usepackage{multirow}
\usepackage{caption}
\usepackage{lipsum}
\usepackage[utf8]{inputenc}

\begin{document}
 \title{Probing Star-Forming Properties via ALMA Observations of Massive Protocluster IRAS 15596-5301}

 \volnopage{ {\bf 20XX} Vol.\ {\bf X} No. {\bf XX}, 000--000}
   \setcounter{page}{1}
   \author{Faxian Chang \inst{1,2} 
   \and Mengyao Tang \inst{2}\footnote{\url{mengyao\_tang@yeah.net}}
   \and Tie Liu \inst{3}
   \and Luis A. Zapata \inst{4}
   \and Dongting Yang \inst{5}  
   \and Yaping Peng \inst{1}
   \and Chao Zhang \inst{6}
   \and  Fengwei Xu \inst{7,8}
   \and Y. H. Chen\inst{2,1}
   \and Shujie Li\inst{2}
    \and Meng Ruan\inst{2}
   }

   \institute{    
    Department of Physics, Faculty of Science, Kunming University of Science and Technology, Kunming 650500, China\\
    \and
   Institute of Astrophysics, Chuxiong Normal University, Chuxiong 675000, China\\
   \and
   Shanghai Astronomical Observatory, Chinese Academy of Sciences, Shanghai 200030, China\\
   \and
   Instituto de Radioastronom\'ia y Astrof\'isica, Universidad Nacional Aut\'onoma de M\'exico, P.O. Box 3-72, 58090, Morelia, Michoac\'an, M\'exico\\
   \and
   School of Physics and Astronomy, Yunnan University, Kunming, 650091, China\\
   \and
   Institute of Astronomy and Astrophysics, School of Mathematics and Physics, Anqing Normal University, Anqing, China\\
   \and
   Kavli Institute for Astronomy and Astrophysics, Peking University, Beijing 100871, China\\
   \and
   Department of Astronomy, School of Physics, Peking University, Beijing, 100871, China\\
\vs \no
   {\small Received 20XX Month Day; accepted 20XX Month Day}
}

   \date{Received September 15, 1996; accepted March 16, 1997}

\abstract{To deepen our understanding of star-forming properties, we studied a massive protocluster IRAS 15596-5301 using ALMA 870 $\mu$m~and 3 mm data. High-resolution 870 $\mu$m~data reveal 34 dense cores, including 3 hot molecular cores, with subsequent line surveys detecting 22 molecular species toward them. Two velocity components (I15596-red/I15596-blue) were found in the averaged H$^{13}$CO$^{+}$(1–0) spectrum, and two filaments were identified from velocity-resolved integrated intensity maps. A spatial overlap between the two filaments was observed, and this overlapping region exhibits a distinct bridge-shaped feature in the position–velocity diagram constructed along the entire filamentary structures. 
Combined with the reduced H$^{13}$CO$^{+}$/HCO$^{+}$ ratio in the overlapping region and the three-dimensional position-position-velocity cube data, we conclude that a non-head-on collision occurs between the edges of the two filamentary structures in IRAS 15596-5301. Cluster analysis demonstrates that clusters located in the collision region host more evolved chemical rich dense cores than their counterparts in other regions. Our results thus indicate that star formation in I15596 is triggered or accelerated by a mild non-head-on collision between two filaments.
} 

\authorrunning{Chang et al. }            
\titlerunning{Properties of IRAS 15596-5301} 
\maketitle

\section{Introduction} \label{intro}

Based on the current research achievements in star formation, a consensus can be obtained that star formation originates in cold, dark molecular clouds \cite[][]{McKee2007ARA&A..45..565M}.
These clouds represent dense regions of the interstellar medium (ISM), predominantly composed of molecular gas and serving as its primary structural component \cite[][]{Rice2016ApJ...822...52R,Miville-Deschenes2017ApJ...834...57M}. 
A critical density threshold must be surpassed in molecular clouds to initiate star formation, whereupon gravitational forces overcome the supporting pressures, including thermal, turbulent, and magnetic forces, triggering gravitational collapse \cite[][]{Zinnecker2007ARA&A..45..481Z}. 
However, relying solely on the self-gravity of the molecular cloud usually makes it difficult to achieve such a high-density state within a reasonable timescale. Therefore, external triggering mechanisms are required to rapidly increase the density of the molecular cloud \cite[][]{McKee2007ARA&A..45..565M}.

Cloud-cloud collisions (CCCs), as one of the important triggering mechanisms \cite[][]{Stone1970ApJ...159..293S,Gilden1984ApJ...279..335G,Torii2015ApJ...806....7T,Issac2020MNRAS.499.3620I}, produce multiple physical effects through violent interactions between colliding molecular clouds. When two molecular clouds collide with each other, strong shock waves and turbulence are generated at the collision interface, resulting in severe compression of the gas and a rapid increase in local density. This dramatic enhancement amplifies self-gravity, establishing the essential conditions for gravitational collapse, so to trigger star formation processes \cite[][]{Habe1992PASJ...44..203H,Anathpindika2010MNRAS.405.1431A,Takahira2014ApJ...792...63T}. At the same time, the high turbulence environment generated by collisions will promote further gas fragmentation, forming more dense cores \cite[][]{Larson1981MNRAS.194..809L,Elmegreen2002ApJ...577..206E,Mac2004RvMP...76..125M}. In addition, the intense turbulent motions form complex networks of filamentary structures (hereafter filaments) within molecular clouds \cite[][]{Inoue2013ApJ...774L..31I,Inoue2018PASJ...70S..53I,Yamada2022MNRAS.515.1012Y}.
These filaments play a key role in star formation by channeling material accumulation and fostering the necessary conditions for subsequent gravitational collapse \cite[e.g.][]{Myers2009ApJ...700.1609M,Liu2021MNRAS.505.2801L,Yang2023ApJ...953...40Y,Sen2024ApJ...967..151S}.

CCCs impose a persistent, systematic impact: beyond initial density enhancement, continuous turbulence regulation progressively reshapes molecular cloud substructure. This dynamic evolution governs both star formation efficiency and stellar spatial distributions. 
Consequently, CCCs not only initiate star formation but also further affect its mass distribution through altering the kinematics of the molecular clouds \cite[][]{Balfour2015MNRAS.453.2471B,Wu2017ApJ...841...88W,Wu2020ApJ...891..168W}.
Consequently, investigating filamentary molecular clouds and CCCs in star-forming regions advances our understanding of both star formation mechanisms and their triggering processes.

IRAS 15596-5301 (hereafter I15596) is a massive protocluster located at a distance of $4.21^{+0.41}_{-0.40}$\ kpc \cite[][]{Reid2019ApJ...885..131R}, which hosts dense cores with different properties (including both low-mass and high-mass cores, \citealt{Xu2024ApJS..270....9X}), and several chemically rich molecular cores have been reported in this region \cite[][]{Li2024MNRAS.533.1583L,Tang2024ApJS..275...25T}. 
\cite{Zhou2022MNRAS.514.6038Z} reported the presence of a filamentary molecular cloud structure in this region, accompanied by multi-velocity components and a velocity gradient distributed across the entire region. \cite{Cyganowski2008AJ....136.2391C} identified an Extended Green Object (EGO) associated with I15596, implying the potential existence of collisionally shock-excited hot gas within the region. Furthermore, a tracer of massive star formation, the 6.7 GHz methanol maser, has been detected to be spatially coincident with the EGO in the region \cite[][]{Green2012MNRAS.420.3108G}. Collectively, these findings demonstrate that I15596 serves as an ideal target for investigating the physical processes underlying massive star formation and molecular cloud collisions.

This paper is structured as follows: Section \ref{obs} details observations and data reductions, Section \ref{res} presents dense core, molecular line and filament identifications, Section \ref{dis} investigates CCC interactions and star-forming properties, and Section \ref{con} summarizes the main findings of this work.

\section{Observations} \label{obs}
\subsection{870 $\mu$m~data}  \label{ALMA}
The observations I15596 were carried out from May 18 to May 20, 2018 (UTC) as part of the ALMA Cycle 5 observations (PI: Tie Liu, Project ID: 2017.1.00545S). The C43-1 configuration was employed for the observations.  A mosaic observation mode was utilized to extend the field of view to be 46$^{\prime\prime}$. I15596 region was completely covered with a small mosaic of seven positions distributed in a Nyquist-sampled grid. Phase, amplitude, and bandpass calibration were performed using J1650-5044 as the phase calibrator, and J1924-2914, J1427-4206, and J1517-2422 for bandpass and flux calibration. 
We performed a total of three rounds of phase self-calibration and one round of amplitude self-calibration using the Common Astronomy Software Applications (CASA) package (version 5.6; \citealt{2022PASP..134k4501C}) to improve the signal-to-noise ratio (SNR) of the images. 
During the imaging process, we set  the deconvolution as ``\texttt{hogbom}'' with a weighting parameter of ``\texttt{briggs}'', utilizing a \texttt{robust} parameter of 0.5 to balance the sensitivity and resolution of final images. A primary beam correction was also implemented. The resulting angular resolution is about 0.8$^{\prime\prime}$, corresponding to a spatial scale of 0.016 pc. 
The largest angular scales (LAS) of observations were determined to be 7.2$^{\prime\prime}$.
The continuum image of 870 $\mu$m is presented as black contours in Figure \ref{fig:fig1} (a).

To cover molecular transitions, the spectral data were obtained in four spectral windows  (SPWs 31, 29, 27, and 25) in Band 7. 
The central frequencies for SPW 25 and SPW 27 are 354.4 GHz and 356.7 GHz, respectively, with a channel spectral resolution of 0.244 MHz. 
The central frequencies for SPW 29 and SPW 31 are 345.1 GHz and 343.2 GHz, respectively, with a channel spectral resolution of 0.976 MHz. 
Each spectral window comprises a total of 1920 channels, resulting in a narrow bandwidth of 468.75 MHz for SPW 25 and SPW 27, and a broad bandwidth 1875 MHz for SPW 29 and SPW 31. 
The detailed information on observations and data reductions for 870 $\mu$m~data can be found in \cite{Tang2024ApJS..275...25T} and \cite{Xu2024ApJS..270....9X}.

\subsection{3 mm data} \label{ATOMS}
The ALMA Band 3 data (Project ID: 2019.1.00685.S; PI: Tie Liu) of I15596 were also included in this work.
The data are part of the ALMA Three-Millimeter Observations of Massive Star-forming Regions (ATOMS) survey, which targets 146 clumps identified by the Infrared Astronomical Satellite (IRAS) \cite[][]{Liu2020MNRAS.496.2790L,Liu2020MNRAS.496.2821L,Liu2021MNRAS.505.2801L}. 
Calibration and imaging were also performed using the CASA software package (version 5.6; \citealt{McMullin2007ASPC..376..127M}). 
The data from the Atacama Compact 7 m Array (ACA; Morita Array) and 12-meter array were calibrated separately, after which the visibility data from both configurations were combined and imaged using CASA.

The SPWs 1 to 6 in the lower sideband have a bandwidth of 58.59 MHz, covering dense gas tracers such as transitions of HCO$^{\rm +}$(1-0), H$^{\rm 13}$CO$^{\rm +}$(1-0), HCN(1-0), and H$^{\rm 13}$CN(1-0), shock tracer SiO(2-1) and photodissociation region tracer CCH(1-0). 
The wider SPWs 7 and 8 in the upper sideband span a frequency range of 97.530 GHz$-$101.341 GHz, each with a bandwidth of 1875 MHz, used for continuum emission and line surveys. 
The channel spectral resolution is $\sim$0.061 MHz, $\sim$0.031 MHz, and $\sim$0.488 MHz for SPWs 1 to 4, SPWs 5 and 6, and SPWs 7 and 8, respectively \cite[][]{Liu2021MNRAS.505.2801L}. 

\begin{figure*}
	\includegraphics[angle=0, width=0.9\textwidth]{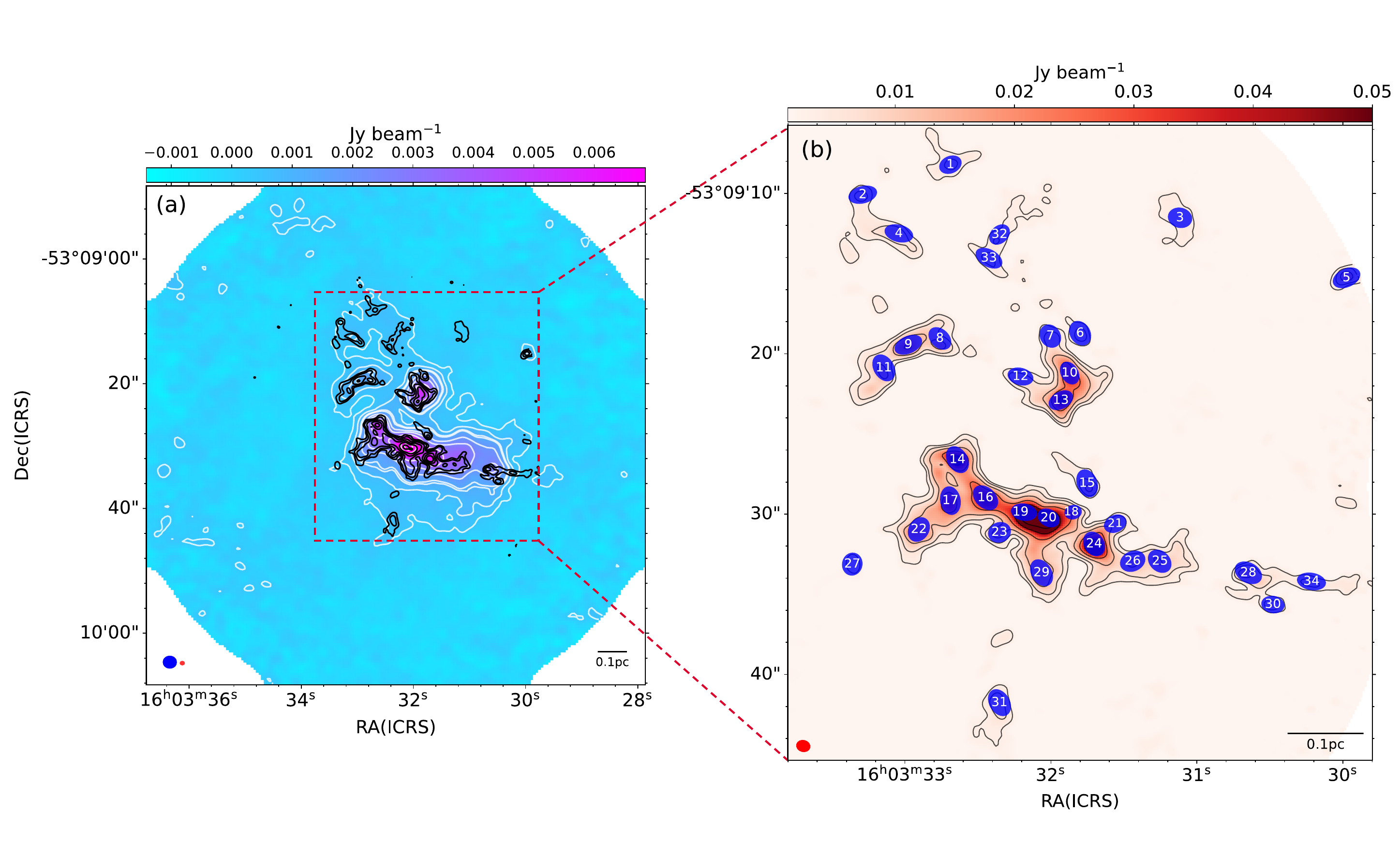}
	\caption{Panel (a): continuum emission at 3 mm wavelenth of I15596 in color scale and with white contours, overlaid with the 870$\mu$m~continuum emission map with the black contours. 
		The white contour levels are [3, 6, 9, 10, 15, 20, 30, 70, 90\%] of peak intensity (13.1 mJy beam$^{-1}$). 
		The black contours levels are [6$\sigma$, 12$\sigma$, 24$\sigma$, 48$\sigma$, 96$\sigma$] ($\sigma_{rms}$ = 0.5 mJy beam$^{-1}$). Panel (b): a zoomed-in view of distribution of identified dense cores in 870 $\mu$m~map. The blue ellipses with numbers denote identified dense cores and their names. The beam sizes of 870 $\mu$m~is shown on the lower left corner with red ellipse, respectively. A 0.1 pc scale bar is shown at the bottom right corner of each panel.}
	\label{fig:fig1}
\end{figure*}

The visibility data from the ACA and 12-m array were jointly cleaned using the task ``\texttt{tclean}" in CASA. 
We used natural weighting and a multiscale deconvolver to optimize sensitivity and image quality. 
All images underwent primary beam correction. The continuum image achieved a typical 1 rms noise ($\sigma$) of approximately 0.2 mJy in a synthesized beam of about 2.2$^{\prime\prime}$.
The continuum image of 3 mm is presented as color background and white contours in Figure \ref{fig:fig1} (a).

\section{Results} \label{res}

\subsection{Dense Cores\label{subsec:dense core}}

\subsubsection{Core Identification\label{subsubsec:parameters}}
We identified dense cores from 870 $\mu$m~continuum data by combining 2D Gaussian fitting with the Astrodendrogram algorithm\footnote{\url{http://www.dendrograms.org/}}. In Astrodendrogram, three key parameters must be set: \texttt{min\_value} (initial value), \texttt{min\_delta} (minimum height of the leaf to consider it independent) and \texttt{min\_npix} (minimum number of pixels of the leaf to consider it independent). We set \texttt{min\_value} = 6$\sigma$ ($\sigma$ = 0.5 mJy beam$^{-1}$ is the root-mean-square noise), \texttt{min\_delta} = $\sigma$, and \texttt{min\_npix} = FWHM (0.80$^{\prime\prime}\approx6.15$~pix at 870 $\mu$m).
Given that the leaf structures identified by the Astrodendrogram algorithm inherently filter out continuum emissions of branches and trunks, we utilize it solely for rapidly locating continuum peak positions of dense cores. Subsequent two-dimensional Gaussian fitting is applied to these positions to determine key physical parameters of dense cores, including size, position angle (PA), integrated flux density, and peak intensity.
The basic parameters of dense cores are listed in Table \ref{tab:Gaussfit}.

A total of 34 dense cores were identified, and the locations of dense cores are denoted by blue ellipses (the cores are labelled with numbers) in Figure \ref{fig:fig1} (b). 
Based on the obtained parameters, the masses of dense cores can be estimated using the following equation \cite[][]{Hildebrand1983QJRAS..24..267H,Liu2021MNRAS.505.2801L}:

\begin{equation}
	\label{eq:core_mass}
	M_{\text{core}} = \frac{D^2 S_\nu \eta}{\kappa_\nu B_\nu (T_{\rm d})},
\end{equation}
where $D$ is the distance, $S\rm _\nu$ is the integrated flux density from 870 $\mu$m~continuum, $\eta=100$ is the gas-to-dust ratio \cite[][]{Lis1991ApJ...380..429L, Hasegawa1992ApJS...82..167H}, $\kappa_\nu=$ 1.89 cm$^2$ g$^{-1}$ is the dust opacity at 870 $\mu$m~\cite[][]{Ossenkopf1994A&A...291..943O}, and $B_\nu (T_{\rm d})$ is the Planck function at the temperature of $T_{\rm d}$ ($T_{\rm d}$ of core 14,19 and 20 used is the excitation temperature of CH$_3$OCHO molecule\footnote{CH$_3$OCHO exhibits the largest number of optically thin transition lines within our observational bandwidth, spanning a range of upper energy levels (E$_{u}$= 80–589 K). These lines provide a robust Boltzmann population distribution, enabling accurate estimation of the excitation temperature. Thus, CH$_3$OCHO serves as a reliable tracer of gas temperature and has been extensively utilized in previous studies \cite[e.g.][]{Xu2024ApJS..270....9X,Li2024MNRAS.533.1583L,Tang2025RAA....25i5020T}.}. 
The $T_{\rm d}$ of the other 31 cores uses average clump temperature $T_{d}$=28.5 K,  \mbox{\citealt{Urquhart2018MNRAS.473.1059U}}). 
The mass estimates are listed in Table \ref{tab:Gaussfit}. 

Assuming optically thin emission from the dust, the column density of H$ _{2}$ can be derived using the following equation \cite[][]{Frau2010ApJ...723.1665F}:
\begin{equation}
	\textit{N(H$ _{2} $)}=\frac{{S_\nu}{\eta}}{\mu
		{m_{H}{\Omega}{\kappa_{\nu}{B_{\nu}(T_d)}}}},
\end{equation}\\
where $\Omega$ is the solid angle subtended by the source, $\mu$ =2.8 is the mean molecular weight of the molecular cloud \cite[][]{Kauffmann2008A&A...487..993K}, and {m$_{H}$} is the mass of a hydrogen atom. The $N_{\rm H_2}$ estimates are listed in Table \ref{tab:Gaussfit}.

\begin{table*}
	\caption{Cores parameters}
	\label{tab:Gaussfit}
	\tiny
	\begin{tabular}{lccccccccccc}  
		\hline      
		Core name & RA & DEC & Source size & PA & Integrated flux & Peak intensity & $T^{a}$ &Mass&$N_{\rm H_2}$ &Catalog\\
		&     &    & ($^{\prime\prime}\times^{\prime\prime}$)  & ($^{\circ}$) & (mJy) & (mJy beam$^{-1}$) &K&($\rm M_{\odot}$) & $\rm cm^{-2}$ &\\
		\hline
		core1 & 16:03:32.681 & $-$53.09.08.155& 2.4$\times$1.3 & 111.0 & 41.6(6.5) & 6.4(0.9) &28.5(5.0)&2.6(0.5)&7.2(1.1) E+22 &Dust&\\
		core2 & 16:03:33.299 & $-$53.09.10.417 & 1.6$\times$0.3 & 171.0 & 25.2(4.7) & 7.2(1.1) & 28.5(5.0) &1.6(0.3)&2.8(0.5) E+23 &Dust&\\
		core3 &16:03:31.131 & $-$53.09.11.759 & 3.7$\times$0.4 & 19.1 & 69.2(7.5) & 4.4(0.4) &28.5(5.0) &4.3(0.6)&2.5(0.3) E+23 &Dust&\\
		core4 & 16:03:33.078 & $-$53.09.12.614 & 2.9$\times$0.4 & 58.0  &56.8(7.0) & 10.1(1.1) & 28.5(5.0) &3.5(0.5)&2.6(0.3) E+23 &Dust&\\
		core5 & 16:03:29.976 & $-$53.09.15.163 & 0.6$\times$0.2 & 108.2 & 23.7(0.3)& 18.4(0.2) & 28.5(5.0) &1.5(0.1)&1.1(0.01) E+24 &Dust& \\
		core6 & 16:03:31.801 & $-$53.09.18.803 & 1.1$\times$0.8 & 22.0  & 20.8(2.3)   & 8.2(0.7) &28.5(5.0)& 1.3(0.2)&1.3(0.1) E+23 &Dust&\\
		core7 & 16:03:31.988 & $-$53.09.19.232 & 1.6$\times$0.8  & 20.9 & 32.9(2.5) & 9.2(0.6) &28.5(5.0)&2.1(0.2) &1.4(0.1) E+23 &Dust& \\
		core8 & 16:03:32.779 & $-$53.09.19.359 & 1.6$\times$0.7 & 68.9   & 45.7(2.6) & 15.1(0.7) & 28.5(5.0) &2.9(0.3)& 2.2(0.1) E+23 &Dust&\\
		core9 & 16:03:32.980 & $-$53.09.19.512 & 1.8$\times$0.8 & 116.4 & 91.2(4.5) & 24.3(1.0) &28.5(5.0) & 5.7(0.6)&3.4(0.2) E+23 &Dust&\\
		core10 & 16:03:31.857 & $-$53.09.21.487 & 2.7$\times$0.8 & 29.1 & 189.2(6.0) & 34.2(0.9) &28.5(5.0)&11.8(1.1)&4.7(0.1) E+23 &Dust& \\
		core11& 16:03:33.213 & $-$53.09.21.605 & 3.4$\times$1.7 & 162.3 & 100.7(8.5) & 8.8(0.7)  &28.5(5.0) &6.3(0.8)& 9.4(0.8) E+22 &Dust&\\
		core12 & 16:03:32.174 & $-$53.09.21.715 & 2.1$\times$0.5 & 67.1 & 34.1(5.0) & 10.2(1.2) &28.5(5.0) &2.1(0.4)&1.7(0.3) E+23 & Dust&\\
		core13 & 16:03:31.923 & $-$53.09.22.844 & 1.2$\times$0.7 & 146.8 & 131.1(2.7)  & 51.8(0.8) &28.5(5.0) &8.2(0.8)& 8.4(0.2) E+23 &Dust&\\
		core14$^{\ast}$ & 16:03:32.645 & $-$53.09.26.667 & 1.3$\times$0.9 & 58.0 & 167.0(19.0) &54.0(4.7) &80.5(3.4)&3.0(0.4)&7.7(0.9) E+23 &CH$_3$OCHO\\
		core15 & 16:03:31.744 & $-$53.09.28.299 & 0.7$\times$0.4 & 21.0 & 20.9(3.0) & 14.0(1.3) &28.5(5.0)& 1.3(0.2) &4.0(0.6) E+23 &Dust&\\
		core16 & 16:03:32.452 & $-$53.09.29.073 & 2.1$\times$1.3 & 60.1 &  251.0(16.0) & 42.2(2.3) &28.5(5.0) &15.7(1.7)&5.0(0.3) E+23 &Dust& \\
		core17 & 16:03:32.697 & $-$53.09.29.333 & 1.7$\times$1.1 & 171.9 & 151.1(7.4) & 34.5(1.4) &28.5(5.0) & 9.4(1.0)&4.4(0.2) E+23 &Dust&\\
		core18 & 16:03:31.904 & $-$53.09.30.041 & 1.3$\times$0.7 & 85.2 & 82.2(0.3) & 29.9(0.1) &28.5(5.0) & 5.1(0.5)&4.9(0.02) E+23 &Dust&\\
		core19$^{\ast}$ & 16:03:32.137 & $-$53.09.30.143 & 0.9$\times$0.7 & 83.8 & 400.2(1.1) & 191.6(0.4) &134.5(16.5) & 4.1(0.4)&3.4(0.01) E+24 &CH$_3$OCHO\\
		core20$^{\ast}$ & 16:03:32.035 & $-$53.09.30.495 & 1.0$\times$0.7 & 92.58 & 333.3(0.2) & 149.0(0.1) &85.2(11.3) &5.6(0.5)&2.6(0.01) E+24 &CH$_3$OCHO\\
		core21 & 16:03:31.554 & $-$53.09.30.624 & 1.3$\times$0.6 &129.2 & 28.5(0.6)  & 11.2(0.2)&28.5(5.0)&1.8(0.2)&2.0(0.04) E+23 &Dust& \\
		core22 & 16:03:32.925 & $-$53.09.31.159 & 1.6$\times$1.2 & 100.7 & 88.5(0.8) & 20.8(0.2) &28.5(5.0) &5.5(0.5)&2.5(0.02) E+23 &Dust& \\
		core23 & 16:03:32.352 & $-$53.09.31.144 & 1.2$\times$0.7 & 117.5 & 39.0(0.4) & 16.0(0.1) &28.5(5.0)& 2.4(0.2)&2.5(0.03) E+23 &Dust&\\
		core24 & 16:03:31.706 & $-$53.09.32.057 & 0.8$\times$0.5 & 63.7 & 164.6(0.9) & 95.8(0.4) &28.5(5.0) &10.3(0.9)&2.2(0.01) E+24 &Dust& \\
		core25 &16:03:31.264 & $-$53.09.32.930 & 2.0$\times$1.3 & 43.4 & 71.2(0.1) & 12.6(0.1) &28.5(5.0) &4.4(0.4)& 1.5(0.01) E+23 &Dust&\\
		core26 & 16:03:31.440 & $-$53.09.33.042 & 1.6$\times$0.9 & 109.0 & 49.4(0.1) & 13.6(0.1) &28.5(5.0) & 3.1(0.3)&1.8(0.01) E+23 &Dust&\\
		core27 & 16:03:33.354 & $-$53.09.33.022 & 0.9$\times$0.4 & 4.3& 9.6(0.1) & 5.4(0.1) &28.5(5.0) & 0.6(0.1)&1.4(0.01) E+23 &Dust&\\
		core28 & 16:03:30.674 & $-$53.09.33.690 & 1.1$\times$0.5 & 82.3 & 30.9(0.3) & 15.9(0.1) &28.5(5.0) & 1.9(0.2)&3.0(0.03) E+23 &Dust&\\
		core29 & 16:03:32.069 & $-$53.09.33.792 & 1.8$\times$1.2 & 10.2 & 87.2(0.9)& 17.5(0.2) &28.5(5.0) &5.4(0.5)& 2.2(0.02) E+23 &Dust&\\
		core30& 16:03:30.490 & $-$53.09.35.60 & 0.8$\times$0.3 & 84.3 & 14.1(0.1) & 9.3(0.1) &28.5(5.0) &0.9(0.1)&3.2(0.02) E+23 &Dust& \\
		core31 & 16:03:32.368 & $-$53.09.41.688 & 3.6$\times$0.9 & 15.0 & 66.1(0.8) & 8.7(0.1) &28.5(5.0) &4.12(0.4)& 1.1(0.01) E+23 &Dust&\\
		core32 & 16:03:32.373 & $-$53.09.12.962 & 1.1$\times$0.8 & 135.6 & 16.5(0.1) & 6.5(0.1) &28.5(5.0) &1.0(0.1)& 1.0(0.01) E+23 &Dust&\\
		core33 & 16:03:32.426 & $-$53.09.14.091 & 1.6$\times$0.5& 45.6 & 22.3(0.7) & 8.1(0.2) &28.5(5.0) &1.4(0.1)& 1.5(0.05) E+23 &Dust&\\
		core34 & 16:03:30.225 & $-$53.09.34.309 & 2.2$\times$0.4 & 79.9 & 24.1(0.6) & 7.3(0.1) &28.5(5.0)&1.5(0.1)&1.5(0.04) E+23 & Dust&\\
		\hline
	\end{tabular}
	\begin{tablenotes}
		\item[ ] {$^{a}$:The temperatures ($T$) for cores 14, 19, and 20 are the excitation temperatures of the CH$_3$OCHO molecule. For the other 31 cores, the temperatures used are the cores' dust temperatures. }\\
		\item[ ]{$^{\ast}$: The marker indicates that the molecular cloud core is recognized as a hot molecular core.}\\
	\end{tablenotes}
\end{table*}

\subsubsection{Molecular species in dense cores}
To study the chemical properties of dense cores, a molecular line survey has been performed on the spectra of dense cores.
The spectra of SPWs 29 and 31 at 870 $\mu$m data were extracted from the continuum peak position of each dense core, and were employed for line surveys.
The spectra of SPWs 25 and 27 were excluded because they have only 468.75 MHz bandwidths, and limited emissions can be detected. These line emissions within these two spectral windows are significantly contaminated by outflow wings of HCN(1-0) and HCO$^{+}$(1-0), respectively.
Additionally, after checking 3 mm spectra, the number of species detected in the 3 mm data is fewer than those in the 870 $\mu$m~data, and the species detected are also fully overlapping with those at 870 $\mu$m.
Therefore, we only presented the results of molecular line identification of SPWs 29 and 31 at the 870 $\mu$m~data. 

The MAdrid Data CUBe Analysis (\texttt{MADCUBA}\footnote{\url{https://madcuba.iaa.es}};\citealt{Martin2019A&A...631A.159M,Blanco2023ascl.soft02019B}) package was employed to perform the line identification.
\texttt{MADCUBA} can use the molecular spectroscopic parameters from databases (e.g. Cologne Database for Molecular Spectroscopy \footnote{\url{http://cdms.de}} \cite[][]{Muller2001A&A...370L..49M, Muller2005JMoSt.742..215M,Endres2016JMoSp.327...95E}) , Jet Propulsion Laboratory \footnote{\url{http://spec.jpl.nasa.gov}} \cite[][]{PICKETT1998883}) to identify molecular transitions.
Additionally, \texttt{MADCUBA} also offers an auto-fit function named \texttt{SLIM}, which is capable of generating synthetic spectra under the  local thermodynamic equilibrium (LTE) assumption with five input parameters (excitation temperature $T$, column density $N$, line width $\Delta V_{\rm dec}$, rest velocity $V_{\rm lsr}$, and source size $\theta$), and automatically fits the LTE model to the observed spectra to obtain optimized parameters of each molecular species \cite[][]{Martin2019A&A...631A.159M}.
A total of 22 molecular species were mainly identified within 3 dense cores (core 14, 19, and 20). 
The molecular parameters ($T$, $N$, $\Delta V_{\rm dec}$, and $V_{\rm lsr}$) of detected species in these three dense cores are listed in Table \ref{tab:parameters}. 
The fractional abundance relative to H$ _{2}$, denoted as $f$ = $N$/$N_{\rm H_2}$, are estimated and listed in Table \ref{tab:parameters}. 
The detailed information of detected transitions of these molecular species are listed in Table \ref{tab:linelists} in the Appendix.

\begin{figure*}
	\centering
	\includegraphics[width=0.9\linewidth]{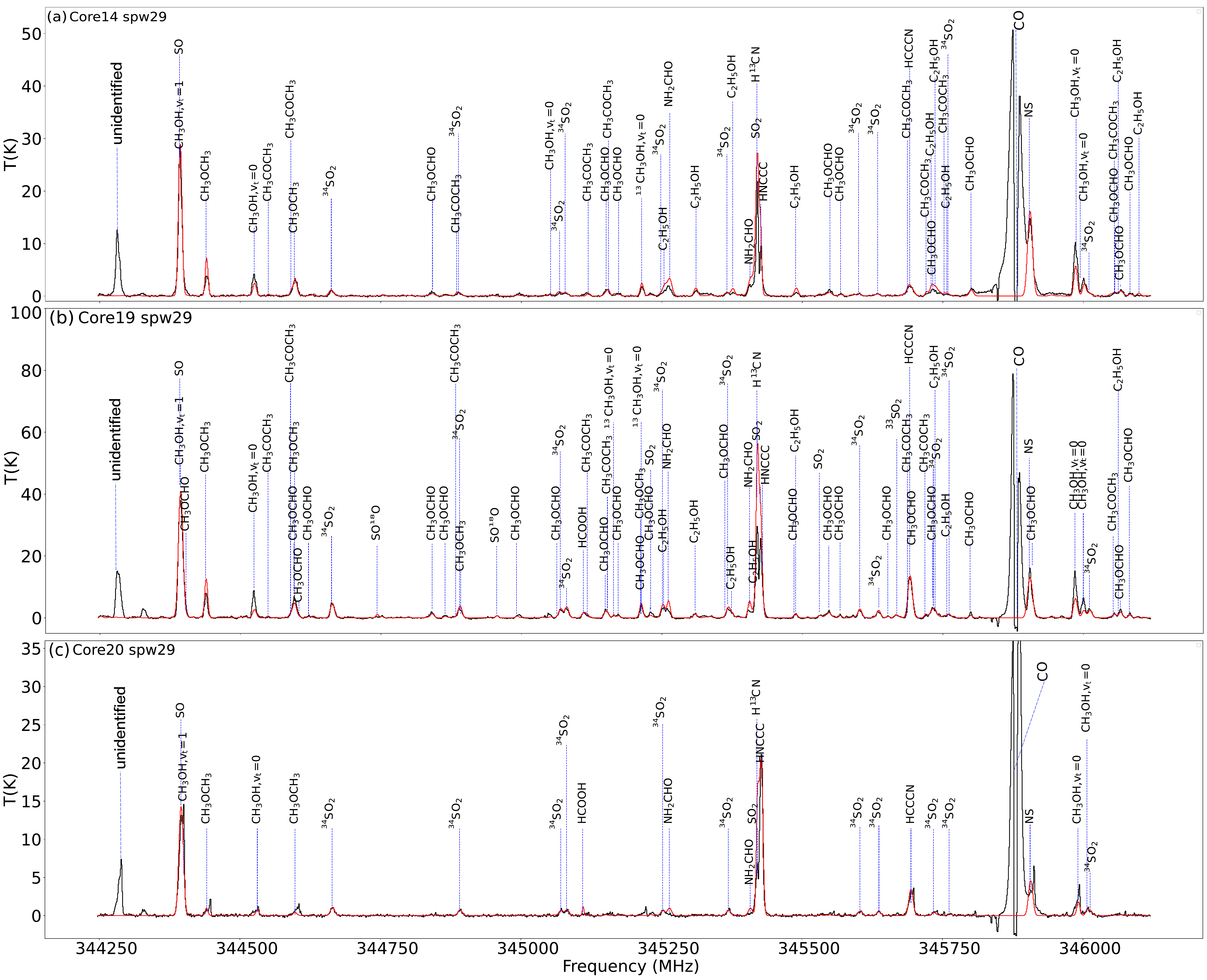}
	\caption{
		Panels (a), (b) and (c): The line survey results of SPW 29, LTE modeling overlaid with observed spectra of SPW 29 at 870$\mu$m~band extracted from the continuum peak position of cores 14, 19 and 20, respectively. The black stepped lines are the observed spectra, while the red curve displays the synthetic spectrum generated by \texttt{MADCUBA} using the optimal-fit parameters. The core names are labeled in the upper left corners. The line names are marked on the plot.}
	\label{fig:fig2}
\end{figure*}

\begin{figure*}
	\centering
	\includegraphics[width=0.9\linewidth]{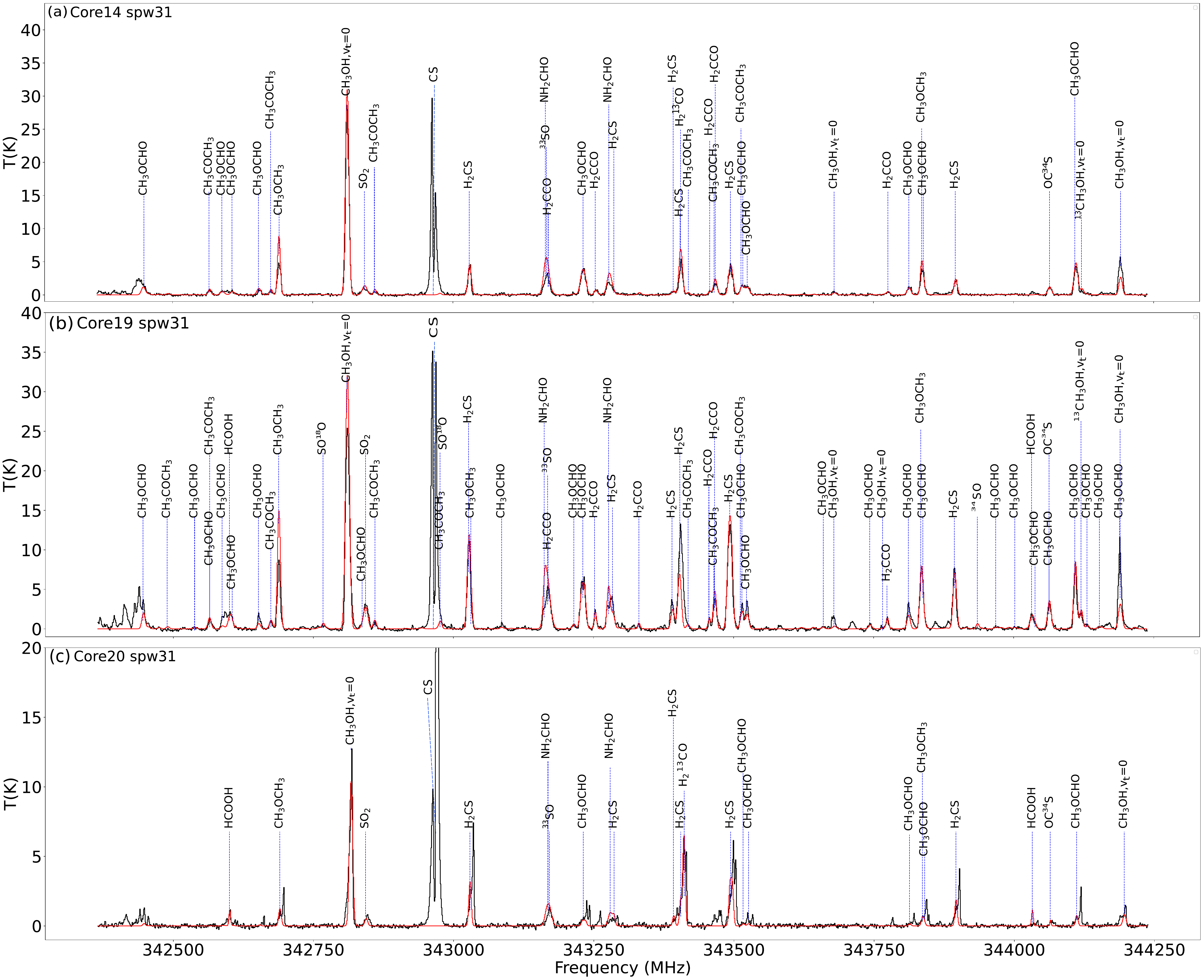}
	\caption{
		Panels (a), (b) and (c): The line survey results of SPW 31, LTE modeling overlaid with observed spectra of SPW 31 at 870 $\mu$m~band extracted from the continuum peak position of cores 14, 19 and 20, respectively. The black stepped lines are the observed spectra, while the red curve displays the synthetic spectrum generated by \texttt{MADCUBA} using the optimal-fit parameters. The core names are labeled in the upper left corners. The line names are marked on the plot.}
	\label{fig:fig3}
\end{figure*}

Based on our results, core 14, 19, and 20 exhibit the most abundant molecular line emissions.
The chemically rich features of these three cores are mainly contributed by the complex organic molecules (COMs), which are commonly detected in hot molecular cores. 
The excitation temperatures of these detected COMs are generally higher than 100 K (see Table~\ref{tab:parameters}), indicating that these three sources are in hot core stage.
Figure \ref{fig:fig2} and \ref{fig:fig3} presents the molecular identification results of cores 14, 19, and 20 in SPW 29 and 31, respectively. 
The transitions are marked by vertical dotted lines, with labeled line names. 
The spectra of other dense cores are omitted for display because only several simple molecules are identified, and they are excluded from further discussion.
Our spectral line survey reveals that the three hot cores exhibit high similarity in their detected species, yet substantial differences in molecular abundances. 
Among them, core 14 shows the highest molecular abundance, followed by core 19, with core 20 being the lowest. 
These suggest that these hot cores may potentially have formed and evolved at distinct epochs within similar circumstellar environments.

\begin{landscape}
	\begin{table}
		\caption{Molecular parameters}
		\label{tab:parameters}
		\tiny
		\begin{tabular}{lcccccccccccccccccc}
			\hline
			\hline
			Molecule &\multicolumn{5}{c}{Core 14} & &\multicolumn{5}{c}{Core 19}  &  &\multicolumn{5}{c}{Core 20} \\
			\cline{2-6}\cline{8-12}\cline{14-18}
			&	$T$ & 	$N$ &	$\Delta V_{\rm dec}$ &	$V_{\rm lsr}$ &	$f$ & &	$T$ & 	$N$ &	$\Delta V_{\rm dec}$ &	$V_{\rm lsr}$&$f$ & &	$T$ & 	$N$ &	$\Delta V_{\rm dec}$ &	$V_{\rm lsr}$ &	$f$ \\
			&	(K) & 	(cm$^{-2}$)  &	(km s$^{-1}$) &	(km s$^{-1}$) &	 &	 & (K) & 	(cm$^{-2}$)  &	(km s$^{-1}$) &	(km s$^{-1}$)&	 &  &	(K) & 	(cm$^{-2}$) &	(km s$^{-1}$) &	(km s$^{-1}$)\\
			\hline
			CH$_{3}$OCHO	              &80.5(3.4)$^{a}$    &2.0(0.1)E+16   &7.1(0.2)	    &$-$71.4(0.1) & 2.6(0.3)E-8 &  &134.5(16.5)$^{d}$    &2.4(0.2)E+16    &5.9(0.2)    &$-$70.4(0.1)  &7.0(0.6)E-9 &  &85.2(11.3)$^{g}$     &2.4(1.5)E+15    &4.6       &$-$71.4&9.2(5.8)E-10\\
			CH$_3$OCH$_3$	      &60.6(2.9)          &1.5(0.1)E+16   &5.0          &$-$71.6      & 1.9(0.3)E-8 &  &134.5$^{d}$          &6.1(1.0)E+16    &5.4         &$-$70.5       &1.8(0.3)E-8 &   &55.6(10.6)           &4.8(1.3)E+15    &5.8(0.7)  &$-$72.0&1.8(0.5)E-9\\
			CH$_3$COCH$_3$	          &90.2(24.4)         &8.7(0.2)E+16   &4.9          &$-$70.3(0.3) & 1.1(0.1)E-7 & &122.3(19.6)          &6.5(1.2)E+15    &5.9         &$-$69.8(0.3)  &1.9(0.3)E-9 &   &$\cdots$             &$\cdots$        &$\cdots$  &$\cdots$&$\cdots$ \\
			H$_2$CS	                  &81.3(5.1)          &2.0(0.1)E+15   &6.0(0.4)	    &$-$72.9(0.2) & 2.6(0.3)E-9 &  &95.3(5.8)            &8.2(0.4)E+15    &6.3(0.3)    &$-$71.3(0.1)  & 2.4(0.1)E-9 &  &104.9(29.8)          &1.5(0.2)E+15    &4.2(0.5)  &$-$72.4&5.8(0.8)E-10\\
			CH$_3$OH           &171.0(8.5)         &1.5(0.1)E+17   &6.9(0.1)	    &$-$71.3(0.1) &1.9(0.3)E-7 &  &233.4(22.9)          &1.2(0.1)E+17    &6.8(0.7)    &$-$70.9(0.3)  & 3.5(0.3)E-8 &  &149.5(38.4)          &2.2(0.2)E+16    &5.6(0.4)  &$-$76.3(0.2)&8.5(0.8)E-9\\
			H$_2$$^{13}$CO            &80.5$^{a}$.        &2.5(0.1)E+14   &6.4(0.3)	    &$-$70.8(0.1) & 3.2(0.4)E-10 &  &$\cdots$             &$\cdots$        &$\cdots$    &$\cdots$      & $\cdots$ & &85.2$^{g}$           &4.7(0.3)E+14    &7.2(0.5)  &$-$5.0(0.2)&1.8(0.1)E-10\\
			H$_2$CCO                  &151.0(13.9)        &9.2(0.3)E+14   &6.1(0.2)     &$-$72.1(0.1) & 1.2(0.1)E-9 & &123.3(7.1)           &2.4(0.1)E+15    &4.6(0.2)    &$-$70.5(0.1)  &7.1(0.3)E-10 &   &$\cdots$.            &$\cdots$        &$\cdots$  &$\cdots$ &$\cdots$\\
			$^{33}$SO                 &137.5$^{b}$        &6.8(0.1)E+14   &9.4(0.8)     &$-$70.0(0.1) &8.8(1.0)E-10 &  &126.2(7.5)           &1.6(5.0)E+15    &10.6(1.0)   &$-$70.5(0.1)  &4.7(14.7)E-10 &   &98.0(19.1)           &2.7(0.1)E+14    &8.3(1.8)  & $-$72.9(0.2)&1.0(0.04)E-10\\
			OC$^{34}$S                &137.5$^{b}$        &1.1(0.6)E+15   &6.9(0.4)     &$-$70.9(0.2) &1.4(0.8)E-9 &  &91.2$^{f}$           &6.1(0.2)E+15    &6.3(0.2)    &$-$70.4(0.1)  &1.8(0.6)E-9 &  &98.6$^{h}$           &3.1(0.8)E+14    &3.0       &$-$71.1(0.5)&1.2(0.3)E-10\\
			HCOOH                     &$\cdots$           &$\cdots$       &$\cdots$     &$\cdots$     & $\cdots$     &  &97.7(18.2)           &2.7(0.6)E+15    &7.5(0.3)    &$-$70.3(0.1)  &7.9(1.8)E-10 &   &126.1(57.7)          &8.8(1.1)E+14    &3.4(0.5)  &$-$69.7(0.3)&3.4(0.4)E-10\\
			SO$_2$                 &155.5(38.7)        &1.5(0.4)E+16   &8.2(0.6)     &$-$71.1(0.3) & 1.9(0.6)E-8 & &141.0(47.7)          &4.6(0.9)E+16    &9.7(0.8)    &$-$71.6(0.3)  &1.3(0.3)E-8 &  &142.4(18.5)          &7.4(2.1)E+15    &8.6(0.7)  &$-$73.0(0.5)&2.8(0.8)E-9\\
			HNCCC                     &153.7$^{c}$        &1.0(0.5)E+14   &3.0          &$-$72.5      & 1.3(0.7)E-10 & &159.1$^{e}$          &7.4(1.4)E+14    &6.0         &$-$71.8       &2.2(0.4)E-10 &  &164.5$^{i}$          &5.5(0.6)E+14    &6.1(0.4)  &$-$73.3(0.2)&2.1(0.2)E-10\\\
			H$^{13}$CN                &153.7$^{c}$        &2.1(0.2)E+14   &5.0          &$-$69.7(0.3) &2.7(0.4)E-10 &  &159.1$^{e}$          &4.8(0.5)E+14    &7.6(1.9)    &$-$69.5(0.5)  &1.4(0.1)E-10 &   &164.5$^{i}$          &1.0(0.7)E+14    &4.2       &$-$70.4&3.8(2.7)E-11\\
			C$_{2}$H$_{5}$OH          &108.9(35.6)        &2.1(0.4)E+16   &11.6(2.4)    &$-$72.0      &2.7(0.6)E-8 &   &116.8(24.7)          &1.6(0.2)E+16    &6.6(0.5)    &$-$69.3(0.2)  & 4.7(0.6)E-9 & &$\cdots$             &$\cdots$        &$\cdots$  &$\cdots$&$\cdots$ \\
			NS                    &129.2(52.3)        &2.0(0.4)E+15   &8.6          &$-$70.8      &2.6(0.6)E-9 &   &91.2$^{f}$           &1.8(0.6)E+15    &8.4         &$-$70.7(1.4)  &5.3(1.8)E-10 &  &98.6$^{h}$           &5.7(2.4)E+14    &8.0       &$-$72.4&2.2(0.9)E-10\\
			HCCCN                     &153.7$^{c}$        &1.7(0.4)E+14   &11.6(0.3)    &$-$71.6(0.1) & 2.2(0.6)E-10 &  &159.1$^{e}$          &1.1(0.1)E+15    &7.6(0.2)    &$-$71.9(0.1)  &3.2(0.3)E-10 &  &164.5$^{i}$          &2.8(0.1)E+14    &8.7(0.5)  & $-$74.7(0.2)&1.0(0.04)E-10\\
			SO                    &137.5$^{b}$        &5.4(0.1)E+15   &6.2(0.2)     &$-$71.8(0.1) &7.0(0.8)E-9 &  &185.7(8.6)           &1.5(0.3)E+16    &7.7(0.4)    &$-$72.2(0.1)  &4.4(0.9)E-9 &   &98.6$^{h}$           &4.4(0.2)E+15    &8.1(0.3)  &$-$74.9(0.2)&1.7(0.8)E-9\\
			$^{34}$SO$_2$             &137.5(48.4)$^{b}$  &1.0(0.2)E+15   &8.4          &$-$70.3(0.6) &1.3(0.3)E-9 &  &91.2(7.5)$^{f}$      &4.9(0.2)E+15    &8.3(0.4)    &$-$71.8(0.2)  &1.4(0.1)E-9 &  &98.6(6.0)$^{h}$      &1.2(0.1)E+15    &7.8(0.3)  &$-$72.9(0.1)&4.6(0.4)E-10\\
			SOO-18                    &$\cdots$           &$\cdots$       &$\cdots$     & $\cdots$    &$\cdots$    &   &110.5(17.2)          &1.7(0.1)E+15    &5.2(1.1)    &$-$72.0(0.1)  &5.0(0.3)E-10 &   &$\cdots$             &$\cdots$        &$\cdots$  &$\cdots$&$\cdots$\\
			$^{33}$SO$_2$             &$\cdots$           &$\cdots$       &$\cdots$     & $\cdots$    &$\cdots$    &  &123.0(59.9)          &8.4(0.9)E+14    &7.4(0.6)    &$-$71.6(0.3)  &2.5(0.3)E-10 &   &$\cdots$             &$\cdots$        &$\cdots$  &$\cdots$&$\cdots$\\
			$^{13}$CH$_3$OH    &92.2(3.0)          &7.7(0.3)E+15   &6.1(0.4)     &$-$71.7(0.2) &1.0(0.1)E-8 &  &146.3(11.8)          &3.4(0.3)E+16    &5.2(0.3)    &$-$69.9(0.1)  &1.0(0.1)E-8 &  &$\cdots$             &$\cdots$        &$\cdots$  &$\cdots$ &$\cdots$\\
			NH$_2$CHO                 &153.7(43.1)$^{c}$  &7.1(1.0)E+14   &7.0          &$-$69.5      &9.2(1.7)E-10 &  &159.1(41.3)$^{e}$    &1.3(0.1)E+15    &9.4(1.1)    &$-$69.5(0.5)  &3.8(0.3)E-10 &  &164.5(74.2)$^{i}$    &2.5(0.6)E+14    &7.7       &$-$71.0&9.6(2.3)E-11\\
			\hline
		\end{tabular}
		\begin{tablenotes}
			\item Notes.The rotation temperatures $T$ and column densities $N$ are model fitting results. The $\Delta V_{\rm dec}$ values is deconvolved line widths. The $V_{\rm lsr}$ is system velocities of the sources. The ‘$\cdots$’ means that no molecular lines were detected.\\
			$^{a, b, d, e f,g ,h, i}$: For some species, the fitting errors of parameters are not given because these species have insufficient transitions to derive parameters. The parameters are derived by using fixed $T$ values from other molecules, which have similar elemental compositions.
		\end{tablenotes}
	\end{table}
\end{landscape}

\subsection{Filaments Identification} \label{fila}
\begin{figure}
	\centering
	\includegraphics[width=0.8\linewidth]{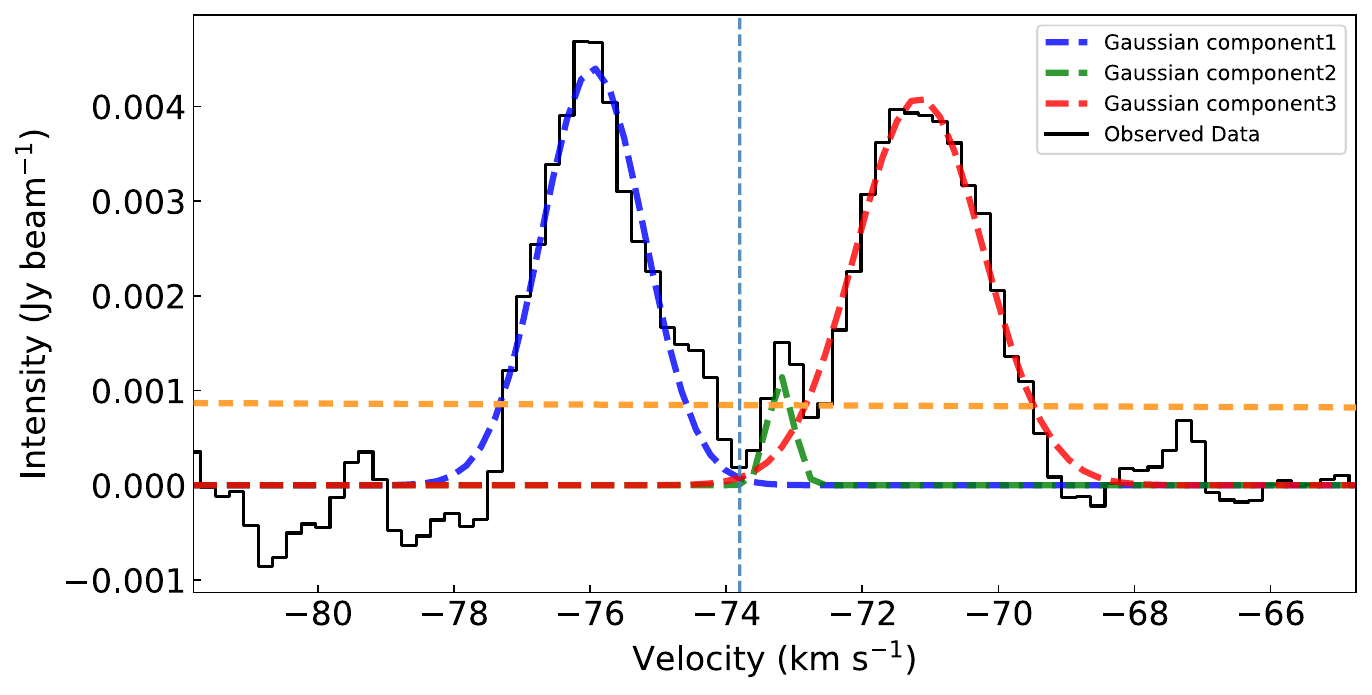}
	\caption{Average spectrum of H$^{13}$CO$^{+}$ (1-0) for the entire surveyed region. The solid black stepped line is the observed H$^{13}$CO$^{+}$ (1-0) spectrum, while the colored dashed profiles represent different Gaussian components. The orange horizontal dashed line denotes the 3$\sigma$$_{\rm ave}$ ($\sigma$$_{\rm ave}$=0.0003 Jy beam$^{-1}$) threshold. Gaussian components with amplitudes below this threshold are excluded from subsequent analysis. The blue vertical dashed line indicates the boundary between two prominent blue-shifted [--78.6, --73.8]  km s$^{-1}$ and red-shifted  [--73.8, --68.5]  km s$^{-1}$ velocity components.
		\label{fig:fig4}}
\end{figure}

\begin{figure*}
	\centering
	\includegraphics[angle=0, width=1\textwidth]{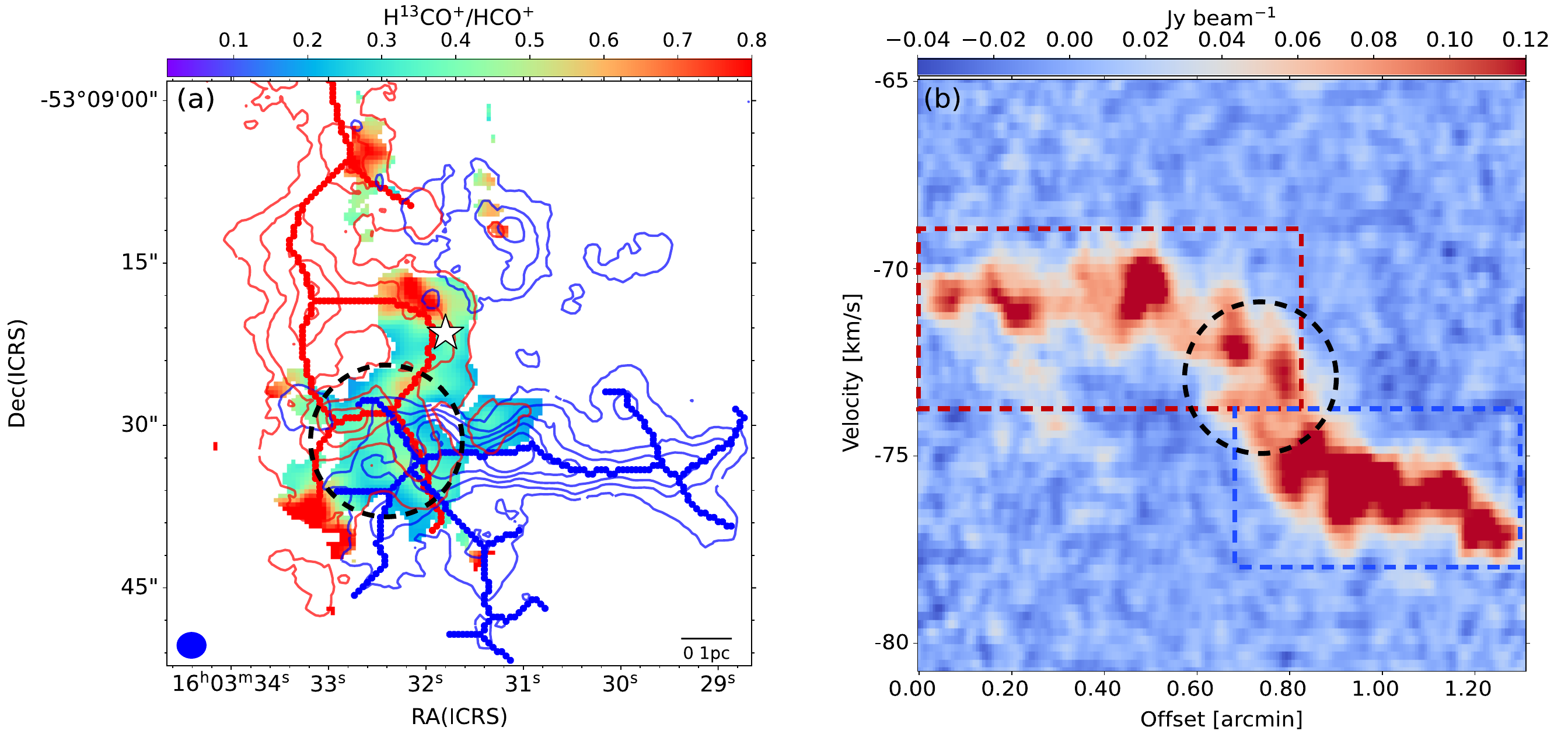}
	\caption{Panel (a): The blue and red contours represent integrated instensity map of blue-shifted (I15596-blue with a velocity range of --78.6 to --73.8 km s$^{-1}$) and red-shifted (I15596-red with a velocity range of --73.8 to --68.5 km s$^{-1}$) component, respectively. 
		The filamentary skeletons identified from H$^{13}$CO$^{+}$(1-0) integrated intensity maps I15596-red and I15596-blue are marked as a red and blue solid bold lines, respectively. The colored background is the H$^{13}$CO$^{+}$/HCO$^{+}$ integrated intensity ratio map, within the velocity range of [--73.5, --72.5] km s$^{-1}$. The dashed black open circle indicate the colliding region. White star denotes bright $Spitzer$ 8 $\mu$m~emission region. The synthesized beam of 3 mm data is shown in a blue ellipse on the lower left corner, and the 0.1 pc scale bar is presented at the bottom right corner. Panel (b): PV diagram of H$^{13}$CO$^{+}$ is generated along the entire filamentary structures. The I15596-red and I15596-blue components are marked by a red and blue dashed boxes, respectively. The black dashed circle denote the bridge-shaped structures.
		\label{fig:fig5}}
\end{figure*}

\begin{figure}
	\centering
	\includegraphics[width=0.9\linewidth]{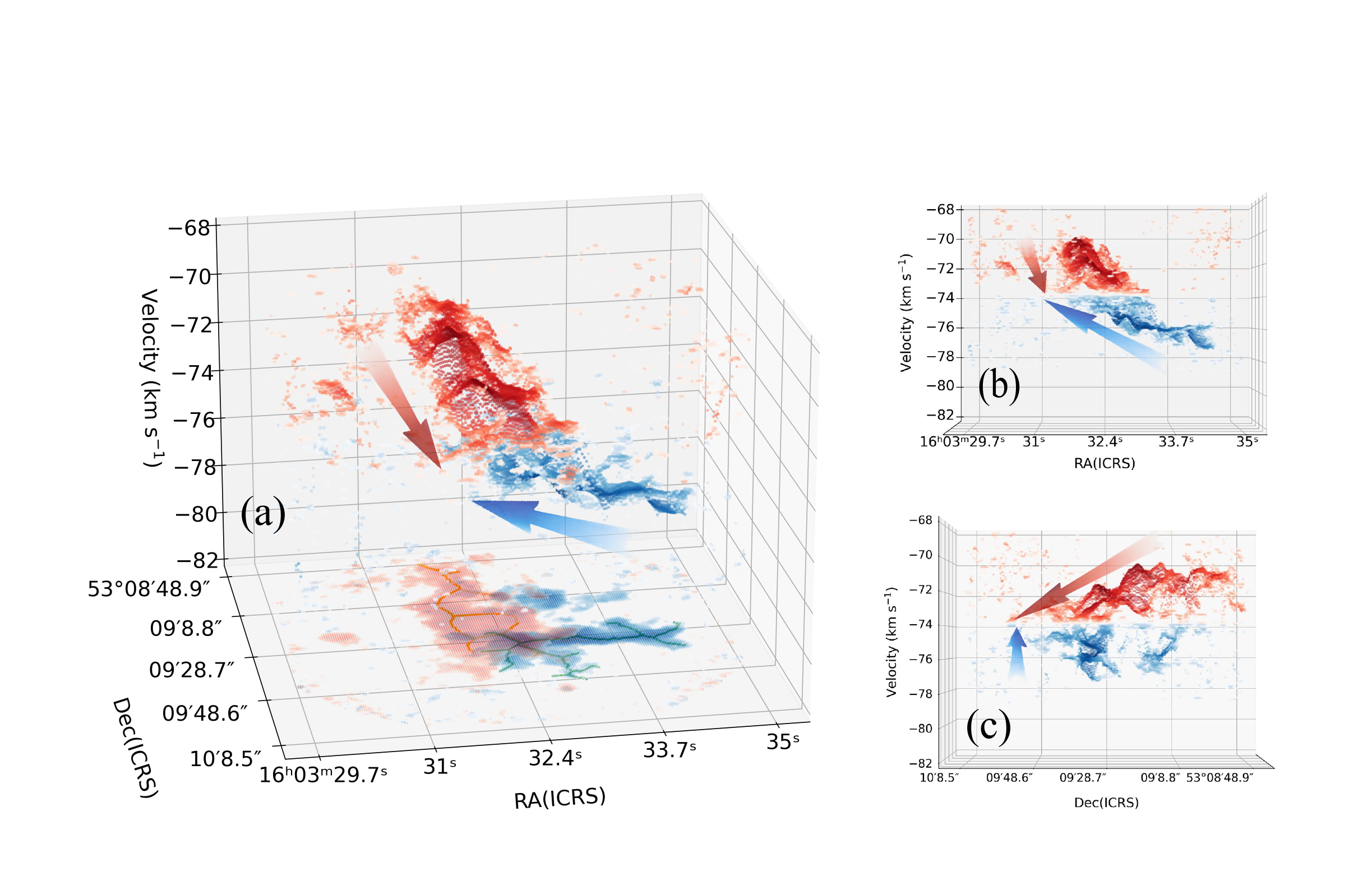}
	\caption{Panel (a): The position-position-velocity (PPV) map of H$^{13}$CO$^{+}$ (1-0) emission for I15596-red and I15596-blue and projection onto a two-dimensional (position-position) plane. Panel (b): Projection onto a position-velocity plane (RA-velocity). Panel (c): Projection onto a position-velocity plane (Dec-velocity). The red and blue scatters represent the intensity of I15596-red and I15596-blue components, respectively. The filamentary skeletons of  I15596-red and I15596-blue are present as yellow and green solid lines on the position-position plane. The red and blue arrows in each panel denote the filamentary motion directions of I15596-red and I15596-blue, respectively.
		\label{fig:fig6}}
\end{figure}

Filaments are widely spread in star-forming regions, and dense cores generally emerge from the converging filaments \cite[e.g.][]{Kumar2020A&A...642A..87K,Liu2021MNRAS.505.2801L,Yang2023ApJ...953...40Y}.
Therefore, studying the properties of filaments is crucial for understanding the core formation and subsequent star formation in molecular clouds. 
The spatial distribution of H$^{13}$CO$^{+}$(1-0) emission is considered to be closely correlated with quiescent dense gas \cite[][]{Shimajiri2017A&A...604A..74S}, making it a good dense gas tracer \cite[][]{Zhou2022MNRAS.514.6038Z}. 
Consequently, this species is frequently employed to identify filamentary structures in star-forming regions \cite[e.g.][]{Zhou2022MNRAS.514.6038Z,Das2024MNRAS.534.3832D,Zhou2024A&A...682A.173Z}.
The present study utilizes 3 mm observational data from the ATOMS project \cite[][]{Liu2020MNRAS.496.2790L}, which includes H$^{13}$CO$^{+}$(1-0) transition. \cite{Zhou2022MNRAS.514.6038Z} employed H$^{13}$CO$^{+}$ (1-0) spectra of ATOMS to generate integrated intensity maps through velocity integration across the whole spectral profile, then using these maps to identify filaments. Subsequently,  a statistical analysis of the properties of filaments was conducted across all ATOMS samples.
While their investigation yields statistically valuable results, their indiscriminate spectral integration, applied without kinematic decomposition, inadvertently obscures velocity-resolved information inherent in the H$^{13}$CO$^{+}$ (1-0) line profiles and the fine-scale internal kinematics.

Figure \ref{fig:fig4} presents the average spectrum of H$^{13}$CO$^{+}$(1-0) for entire surveyed region of I15596. 
Gaussian fittings were performed to identify different H$^{13}$CO$^{+}$(1-0) velocity components. 
From colored dashed lines in Figure \ref{fig:fig4}, a total of 3 velocity components were found, including 2 strong components (peaks 1 and 3) and a weak component (peak 2). 
It should be noted that two weak emission peaks are also visible in Figure \ref{fig:fig4} at velocities of --74.4 and --67.3 km s$^{-1}$. 
	However, after subtracting the three prominent velocity components, the peak intensities of these two components fall below the 3$\sigma_{\rm ave}$ detection threshold and are thus excluded from subsequent analysis.
Our results suggest that the gas traced by H$^{13}$CO$^{+}$(1-0) in this region potentially exhibits discrete structures and distinct kinematic features.
Consequently, to advance understanding of the properties of filamentary structures in I15596, we perform filament identification incorporating velocity component decomposition using H$^{13}$CO$^{+}$(1-0) spectral data.

Based on the average spectrum presented in Figure \ref{fig:fig3}, we performed a velocity-integrated analysis of H$^{13}$CO$^{+}$(1-0) spectra in this region. 
The integration velocity ranges are [--78.6, --73.8] km s$^{-1}$ for blue-shifted component (hereafter, I15596-blue) and [--73.8, --68.5] km s$^{-1}$ for red-shifted component (hereafter, I15596-red). 
The vertical dashed line in Figure \ref{fig:fig3} indicates the division between blue- and red-shifted components.
Due to severe spectral blending between components 2 and 3, coupled with component 2's significantly weaker intensity relative to component 3, we incorporated component 3 directly into the I15596-red integration domain without separate treatment.

Based on velocity component decomposition (Figure \ref{fig:fig4}), we generate H$^{13}$CO$^{+}$(1-0) integrated intensity maps for the I15596-blue and I15596-red components with a 3$\sigma$ ($\sigma$=0.0122 Jy beam$^{-1}$) threshold.
These integrated intensity maps are presented as blue and red contours in Figure \ref{fig:fig5} (a).
We identify filamentary structures using \texttt{FILFINDER} \cite[][]{Koch2015MNRAS.452.3435K}, applying the algorithm separately to each velocity component map (I15596-blue and  I15596-red). 
The resulting filamentary skeletons are shown as solid blue and red lines in Figure \ref{fig:fig5} (a).
A distinct spatial overlap exists between filaments in I15596-blue and I15596-red integrated intensity maps (black dashed circle). 
Relative to the ambient environment, this overlapping region exhibits enhanced gas intensity, manifested as denser emission contours. 
This morphology suggests potential physical interaction between the different kinematic components, which will be discussed in Section \ref{sec: ccc}.

\section{Discussion}\label{dis}
\subsection{Cloud-cloud collision in I15596}\label{sec: ccc}
\begin{figure*}
\includegraphics[angle=0, width=1\textwidth]{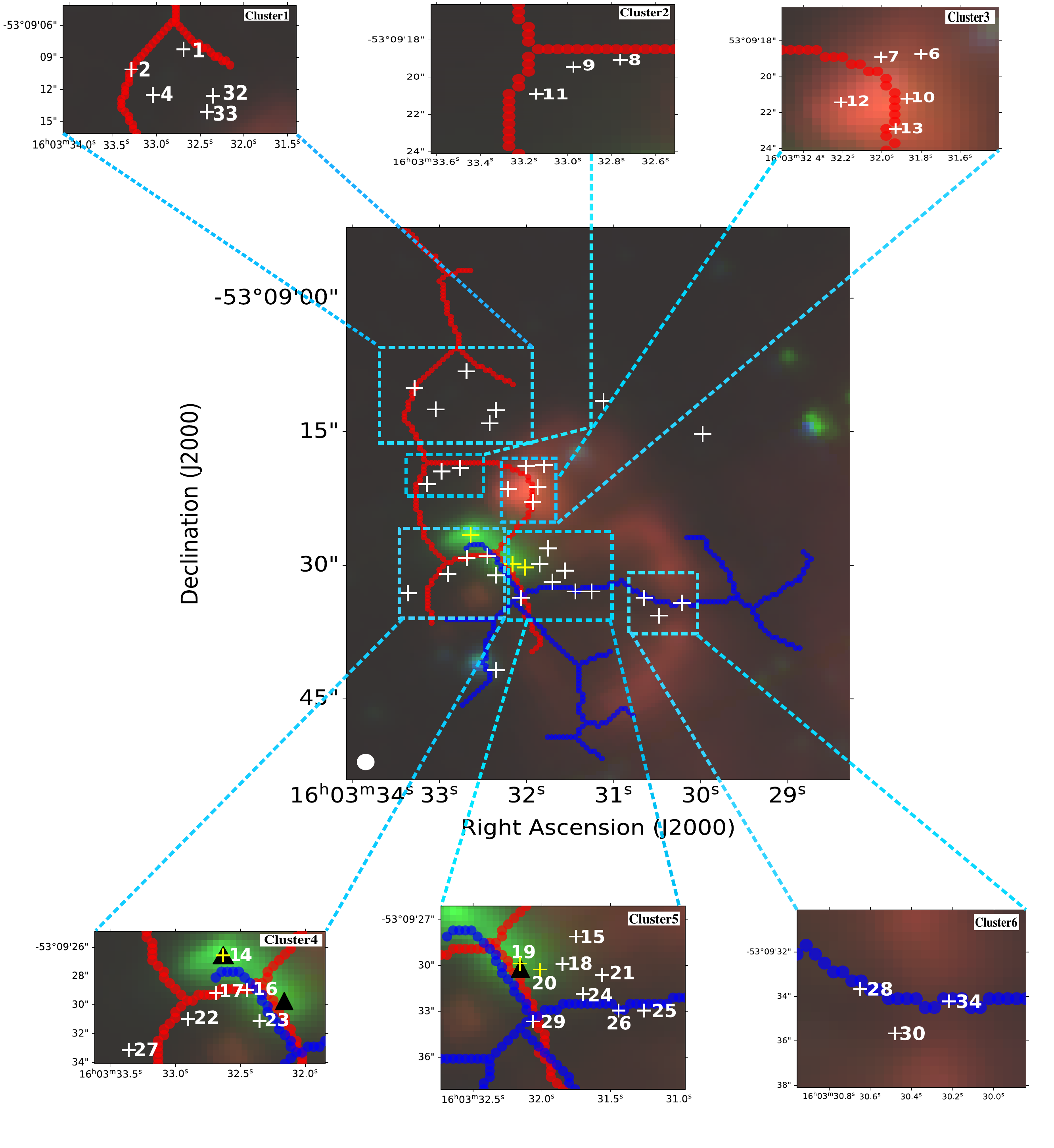}
	\caption{
		The background is an RGB composite image constructed from $Spitzer$ 3.6 $\mu$m (blue), 4.5 $\mu$m (green), and 8 $\mu$m (red) data.
		The filamentary skeletons of I15596-red and I15596-blue are overlaid as red and blue bold solid lines, respectively. The positions of 34 dense cores are marked with colored crosses (yellow crosses for 3 hot cores, and white crosses for other dense cores). Clusters identified by the $K$-means method are enlarged as sub-panels, and the core names are labeled. The names of the clusters are labeled at the upper-right corner in each sub-panel. Two black triangles indicate the positions of the 6.7 GHz CH$_3$OH masers \cite[][]{Green2012MNRAS.420.3108G}. The resolution of the $Spitzer$ observation is shown as a white circle in the bottom-left corner.
		\label{fig:fig7}}
\end{figure*}

As stated at the end of Section \ref{fila}, the filaments identified from I15596-blue and I15596-red seem to interact dynamically with each other.
To further investigate the interactions between them, we generated a position-velocity (PV) diagram as shown in the color scale of Figure \ref{fig:fig5} (b).
PV diagrams are generated along the filamentary structures for entire region. In the right panel of Figure \ref{fig:fig5}, the red and blue rectangles represent the velocity ranges for I15596-red and I15596-blue components, respectively.
As can be seen from the PV diagram, the red- and blue-shifted filamentary structures correspond to distinctly different velocity components, with a maximum velocity difference of approximately 7 km s$^{-1}$ between them. 
The PV diagram reveals a prominent bridge-shaped structure connecting these two velocity components, and the spatial location of this bridge-shaped structure coincides with the overlapping region of I15596-red and I15596-blue filaments (as indicated by a black dashed circle).
Such bridge shaped structures (kinematically bridging distinct velocity components) demonstrate direct evidence of gas interaction and momentum transfer at collision interfaces \cite[][]{Fukui2021PASJ...73S...1F, Berdikhan2025arXiv250414943B}
The observed morphology, connecting I15596-red and I15596-blue, confirms a physical collisional interaction, providing critical insight into molecular cloud evolution and star-formation triggering mechanisms.

Shocks are widely recognized as a crucial tracer of cloud-cloud collisions. 
Observations at 3 mm wavelength cover SiO emission (a well-known shock tracer).
However, prominent outflow motions \cite[][]{Baug2020ApJ...890...44B} in this region cause SiO emission to be strongly concentrated toward outflow lobes, thus precluding its use for tracing low-velocity shocks induced by the collision interface of the filaments. 
For this reason, we have incorporated the integrated intensity ratio map of H$^{13}$CO$^{+}$/HCO$^{+}$ into Figure \ref{fig:fig5}(a). 
As can be seen from Figure \ref{fig:fig5}(a), the overlapping region exhibits a low H$^{13}$CO$^{+}$/HCO$^{+}$ intensity ratio ($\sim$0.4), which is significantly lower than the corresponding values observed in the northern and southeastern regions ($\sim$0.8).
HCO$^{+}$ is highly sensitive to shock excitation and is thus widely employed as a tracer to probe shocked gas in outflows and supernova remnants\mbox{\cite[e.g.][]{Sanchez2013A&A...557A..94S,Tu2024ApJ...966..178T}}. By contrast, H$^{13}$CO$^{+}$ lacks such shock sensitivity. We therefore infer that the enhanced HCO$^{+}$ emission in the overlapping region is likely driven by collision-induced shocks, which consequently gives rise to the observed low H$^{13}$CO$^{+}$/HCO$^{+}$ intensity ratio in this region.
It should be noted that, as revealed by the PV diagram, the collision is concentrated within the velocity range of --73.5 to --72.5 km s$^{-1}$. 
Therefore, we only presented the integrated intensity ratio map of H$^{13}$CO$^{+}$/HCO$^{+}$  within this specific velocity interval, while excluding other non-collisional regions from the analysis.
	
To better visualize their interactions, Figure \ref{fig:fig6}(a) presents a three-dimensional position-position-velocity (PPV) cube, with projections onto the two-dimensional planes overlaid with the filamentary skeletons. 
In the three-dimensional structure of Panel (a), the separation and interaction between the I15596-red and I15596-blue in the PPV space are clearly discernible. 
The arrows indicate the motion directions of the I15596-red and I15596-blue filaments, respectively. 
Panel (b) and (c) of Figure \ref{fig:fig6} present the PPV diagram projected onto position-velocity planes (RA-velocity and Dec-velocity). 
From Panels (b) and (c), the positions of the collision cross-sections of the two filamentary components can be visually identified. 
It is also evident from the panels that the collision between the two filamentary components occurs as a non-head-on impact at the edges of the two filaments.

Figure \ref{fig:fig7} shows an RGB composite image from $Spitzer$ 3.6 $\mu$m (blue), 4.5 $\mu$m (green), and 8 $\mu$m (red) data. As can be seen from the figure, the overlapping region of the filamentary skeletons exhibit prominent 4.5 $\mu$m emission, yet lacks of corresponding 3.6 $\mu$m blue radiation. 
This indicates that the 4.5 $\mu$m emission predominantly originates from molecular line radiation (e.g., H$_2$ and CO) excited by shocks, rather than from infrared dust continuum \mbox{\cite[][]{Reach2006AJ....131.1479R}}. 
Extended infrared excess emission at 4.5 $\mu$m has been detected in EGOs\mbox{\cite[][]{Cyganowski2008AJ....136.2391C}} and supernova remnants\mbox{\cite[][]{Blair2007ApJ...662..998B}}, a phenomenon widely attributed to excitation by outflows or intense shocks. Within the overlapping region, the coexistence of three hot cores, two CH$_{3}$OH masers (see Section\ref{star formation}), and extended 4.5 $\mu$m emission provides compelling evidence for ongoing massive star formation. Therefore, the extended 4.5 $\mu$m emission is most likely excited by outflows launched from the forming massive stars, given that low-velocity cloud-cloud collisions are unlikely to generate sufficient energy to drive such emission.
Multiple observational evidences consistently demonstrate that star formation in I15596 is most vigorous within the collision region, implying that such activity in this region is likely triggered or accelerated by the collision of filamentary molecular clouds. A comprehensive discussion of this underlying mechanism is presented in Section \ref{star formation}.

Notably, other characteristic features of CCCs, such as the complementary spatial morphology (keyhole-shaped structure) between colliding molecular clouds \cite[][]{Berdikhan2025arXiv250414943B}, are absent in the I15596. This absence may arise from the following plausible factors:
\begin{enumerate}
	\item[(1)] The complementary structure primarily arises from collisions between molecular clouds with significant size disparities, where the larger cloud develops a concave indentation that spatially interlocks with the smaller cloud \cite[][]{Fukui2021PASJ...73S...1F}. 
	However, our observations of I15596-red and I15596-blue reveal a similar size between them. 
	Moreover, the collision occurs only within a limited overlapping region, with integrated flux densities at the collision interface measuring 4.98 Jy beam$^{-1}$ and 5.87 Jy beam$^{-1}$ for red- and blue-shifted components, respectively. 
	Accounting for statistical uncertainties of at least 10\%, both the integrated flux densities and spatial sizes of the colliding regions of I15596-red and I15596-blue appear highly comparable. 
	
	\item[(2)] The collision between I15596-red and I15596-blue is not an idealized head-on collision with spherical clouds, and the colliding region is confined to the peripheries of the molecular clouds, which likely explains the partial inconsistency between our observational results and theoretical models.
	
	\item[(3)] Early studies have reported the presence of H$_{\rm II}$ regions in the surveyed region \cite[][]{Garay2002ApJ...579..678G}. However, their insufficient spatial resolution precluded robust comparative analysis with our results. 
		In contrast, previous works \cite[][]{Liu2021MNRAS.505.2801L,Zhang2023MNRAS.520.3245Z} utilizing the same dataset as employed in our work failed to detect any radio recombination lines (RRLs) within this region. 
		Nevertheless, the existence of two associated 6.7 GHz methanol masers \cite[][]{Green2012MNRAS.420.3108G} in the colliding region suggests that ionized gas is most likely confined within a potential ultra-compact (UC) H$_{\rm II}$ region. 
		All of this observational evidence indicates that the collision is not intense (with a maximum velocity difference of only 7 km s$^{-1}$) or that the star-forming activity triggered by the collision has not significantly disrupted the overall structure of the I15596 complex.
\end{enumerate}

\subsection{Star formation triggered by filaments collision in I15596}\label{star formation}
\begin{figure}
	\includegraphics[width=0.9\linewidth]{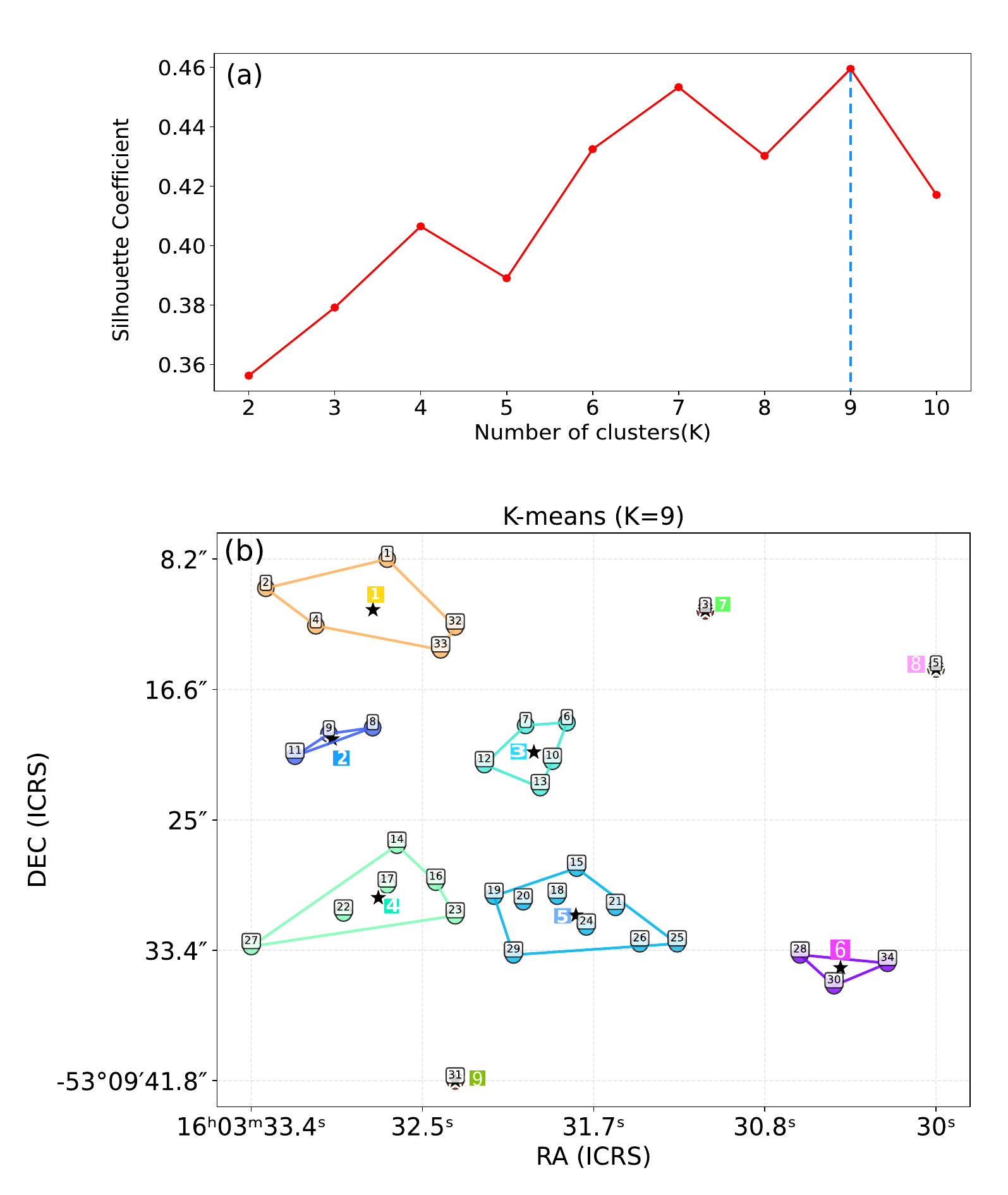}
	\caption{Panel (a): the variation of silhouette coefficient with the number of clusters ($K$). The blue vertical dashed line represents the maximum silhouette coefficient at $K$=9. Panel (b): clusters of I15596 using $K$-means algorithm at $K$=9. Different colors distinguish each cluster, black asterisks indicating cluster centers, and the convex hull depicting cluster boundaries. The cluster number is marked at the center of the cluster.
		\label{fig:fig8}}
\end{figure}

The star and cluster formation mechanism within filamentary molecular clouds constitutes a focal point of astronomical investigations. 
Especially, the CCC dynamic process has been observed in the surveyed region, which is highly related to star and cluster formations.
To deepen our understanding of star formation in I15596, we performed a clustering analysis on the dense cores identified from 870 $\mu$m.
The $K$-means clustering algorithm in the python package  \texttt{scikit-learn}\footnote{\url{http://scikit-learn.org/ stable/modules/clustering.html}}\cite[][]{Pedregosa2011JMLR...12.2825P} was adopted to group the 34 identified continuum dense cores as shown in crosses in Figure \ref{fig:fig7}. 
The $K$-means clustering algorithm requires the expected number of clusters $K$ as an input parameter.
Silhouette analysis was employed to determine the optimal $K$ value by comparing coefficients for $K \in$ [2, 10].
Each candidate $K$ is assigned a silhouette coefficient $\in$ [-1, 1], where higher values indicate superior cluster configuration \cite[][]{Colombo2015MNRAS.454.2067C, Sadaghiani2020A&A...635A...2S}.
Figure \ref{fig:fig8} (a) displays the relationship between silhouette coefficients and $K$, revealing a peak at $K$=9.
Consequently, $K$-means clustering analysis was performed with $K$=9, and the resulting clusters are shown in Figure \ref{fig:fig8} (b).
These 34 cores were grouped into 9 groups, with cores 3, 5, and 31, which were separated into individual groups, respectively. 

It was found that Clusters 4 and 5, located near the collision interface, exhibit the highest number of associated dense cores. 
This observation aligns with the theoretical scenario of collision-triggered core formation, whereby collisions initially generate high-density regions, while turbulence generated by the collisions promotes the fragmentation of dense molecular clouds, thereby forming a larger population of dense cores \cite[][]{Habe1992PASJ...44..203H,Anathpindika2010MNRAS.405.1431A,Inoue2013ApJ...774L..31I,Takahira2014ApJ...792...63T,Inoue2018PASJ...70S..53I}. 
This process establishes optimal initial conditions for subsequent star formation. 
Furthermore, three hot cores were found located close to the overlap regions of filamentary skeletons. 
This aligns with prior studies indicating that collisional feedback serves as a critical mechanism for triggering and enhancing star formation \cite[][]{Whitworth1994MNRAS.268..291W,Takahira2014ApJ...792...63T,Gong2017ApJ...835L..14G,Issac2020MNRAS.499.3620I,Fukui2021PASJ...73S...1F}.
Notably, Clusters 4 and 5 spatially associate with I15596-red and I15596-blue components, respectively. 
Both harbor abundant dense cores and include at least one chemically rich hot core.

Dense cores 14 and 19 are respectively associated with two 6.7 GHz methanol masers \cite{Green2012MNRAS.420.3108G}, indicating that these two cores are typical massive star-forming regions. 
However, it is noteworthy that the masses of these two dense cores derived from our calculations are both below 8 M$_{\odot}$ (as listed in Table \ref{tab:parameters}), implying that their masses have been underestimated. 
CH$_{3}$OCHO rotational temperature was used to estimate the masses of these three cores, substituting a dust temperature of 28.5 K would yield masses that meet the massive core criteria.
This indicates that the UC H$_{\rm II}$ region generated by the central protostar has not yet had sufficient time to heat all the surrounding gas. Only the innermost regions of the dense cores exhibit relatively high temperatures, while the overall average temperature should be lower than the gas temperature derived from CH$_3$OCHO. 
Combining the abundant emission of organic molecules and elevated gas temperatures of detected species, our results suggest the existence of unexpanded UC H$_{\rm II}$ regions within the dense cores 14 and 19.

Our core identification and molecular line detection work across the entire I15596 region reveals a large number of dense cores and molecules. 
However, with the dense cores located out of the collisional region, the chemical composition of cores is rather simple, characterized solely by the emission of a few simple molecules, and no complex organic molecules have been detected in these cores. 
Therefore, synthesizing the results presented throughout this work, we propose the following scenario: two filamentary molecular clouds with a maximum velocity difference of approximately 7 km s$^{-1}$ undergo a collision at their edges. 
Furthermore, the evolutionary stage of dense cores in the colliding region is distinctly later than that in non-colliding regions, indicating that the mild low-velocity collision within I15596 has triggered or accelerated the massive star formation process in this region.

\section{Conclusions} \label{con}
Based on ALMA 870 $\mu$m~and 3 mm observational data, we conducted analyses of the properties of dense cores and filamentary structures in I15596. The primary results of this work are summarized as follows:
\begin{enumerate}
	\item[(1)] A total of 34 dense cores were detected from the 870 $\mu$m~continuum emission. The basic parameters of these cores were estimated and reported.
	
	\item[(2)] We identified three hot molecular cores, in which spectral line surveys detected a total of 22 molecular species. Molecular parameters, including rotation temperature ($T$), column density ($N$), line width ($\Delta V_{\rm dec}$), and system velocity ($V_{\rm lsr}$), were derived by fitting the observed spectra with an LTE model provided by the \texttt{MADCUBA} software package.
	
	\item[(3)] The spatially averaged H$^{13}$CO$^{+}$(1-0) spectrum over the entire region exhibits distinct multiple velocity components. Integrated intensity maps were generated using the two most prominent blue-shifted (with a velocity range of --78.6 to --73.8 km s$^{-1}$) and red-shifted (with a velocity range of --73.8 to --68.5 km s$^{-1}$) components. 
	Based on these velocity-component-resolved integrated intensities, we identified filaments and observed spatial overlaps between adjacent filaments, suggesting potential ongoing interactions.
	
	\item[(4)] 
	Based on the integrated intensity map and position-velocity (PV) diagram constructed along the entire filamentary structures, we found a spatial overlap between two filamentary structures, which were connected by a bridge-shaped feature in the PV space, which represents one of the most prominent signatures of cloud-cloud collisions (CCCs). 
	Relative to the H$^{13}$CO$^{+}$/HCO$^{+}$ intensity ratios measured in other regions, the reduced ratio observed in the overlapping region provides additional confirmation of ongoing collisional interactions. 
	Combining the morphological characteristics of the filament skeletons, a maximum velocity difference of $\sim$7 km s$^{-1}$, the spatial confinement of methanol masers, and the non-detection of recombination lines (RRLs), our results demonstrate that this interaction corresponds to a mild non-head-on collision.
	
	\item[(5)] Using a \texttt{Python}-implemented $K$-means clustering algorithm, we classified 34 dense cores into 9 groups, with six of them identified as clusters. Clusters 4 and 5, proximal to the CCC interface, exhibit the highest dense core density. This supports theoretical models where collisions generate high-density regions, and collision-induced turbulence promotes molecular cloud fragmentation, increasing the number of dense cores. 
	Compared to other regions, the collision region harbors three more evolved hot cores, two of which are associated with methanol masers, indicating massive star formation within the dense cores. Additionally, the underestimated masses of these two cores imply the presence of potential ultra-compact (UC) H$_{\rm II}$ regions within them. Collectively, these results demonstrate that the collision has triggered or accelerated the star formation process in I15596.
\end{enumerate}

\section*{Acknowledgements}
This work has been supported by the National Science Foundation of China (12203011). 
M.Y.T. acknowledge the support by Yunnan province Xingdian talent support program, Yunnan provincial Department of Science and Technology through grant No.202101BA070001-261, Yunnan University Laboratory Open Project, and PhD research startup foundation of Chuxiong Normal University.
T.L by the international partnership program of Chinese Academy of Sciences through grant No.114231KYSB20200009, Shanghai Pujiang Program 20PJ1415500, the science research grants from the China Manned Space Project with no. CMS-CSST-2021-B06.
L.A.Z. acknowledges financial support from CONACyT-280775, UNAM-PAPIIT IN110618, and IN112323 grants, M\'{e}xico.
Y.P. Peng acknowledges support from NSFC through grant No. 12303028.

This paper uses the following ALMA data: ADS/JAO.ALMA\#2017.1.00545.S.
ALMA is a partnership of ESO (representing its member states), NSF (USA) and NINS (Japan), together with NRC (Canada), MOST and ASIAA (Taiwan), and KASI (Republic of Korea), in cooperation with the Republic of Chile. 
The Joint ALMA Observatory is operated by ESO, AUI/NRAO, and NAOJ. 
This research has made use of the NASA/IPAC Infrared Science Archive, which is funded by the National Aeronautics and Space Administration and operated by the California Institute of Technology.

\textbf{Data availability}

The data underlying this article are available in the ALMA archive.

\textbf{software}

This research made use of astropy \cite[][]{2013A&A...558A..33A,2018AJ....156..123A,2022ApJ...935..167A}, MADCUBA\cite[][]{Martin2019A&A...631A.159M} and CASA\cite[][]{2022PASP..134k4501C}.


\appendix

\clearpage

\section{List of all detected transitions} 
In this section, the detailed information, including rest frequency (MHz), transition, line strength ($S_{ij\mu}$$^{2}$), column density of each molecule, and upper-level energy ($\rm E_u$), of detected transitions is listed in Table\,\ref{tab:linelists}. 


\setcounter{table}{0}
\setlength{\tabcolsep}{5pt}
\onecolumn
\begin{longtable}{lccccccccccccccc}
	\caption{Detected transitions}
	\label{tab:linelists}\\
	\hline
	Molecular Name & Frequency & 	Transition & $\rm S_{\rm ij}$$\rm \mu^2$ & 	$\rm Log_{\rm 10}(A_{\rm ij})$ & E$_{\rm u}$ &  \\
	& (MHz) & & (D$^{\rm 2}$) & ($\rm s^{-1}$) & (K)\\
	\endfirsthead
	\caption{continued.}\\
	\hline
	Molecular Name & Frequency & 
	Transition & $\rm S_{\rm ij}$$\rm \mu^2$ & 
	$\rm Log_{\rm 10}(A_{\rm ij})$ & E$_{\rm u}$ &  \\
	& (MHz) & & (D$^{\rm 2}$) & ($\rm s^{-1}$) & (K)\\
	\hline
	\endhead
	\hline
	\endfoot
	\hline
	CH$_3$OCHO v=0   &342366.296	   &30(2,28)-29(2,27) A	&78.02813	&-3.22367	&269.49059&\\
	&342367.6798   &30(3,28)-29(2,27) E	&10.0013	&-4.11586	&269.4967&\\
	&343147.8978	&31(1,30)-30(2,29) E	&11.85124	    &-4.0532	&273.44227&\\
	&343148.0473	&31(2,30)-30(2,29) E	&81.50109	    &-3.2158	&273.44228&\\
	&343148.1688	&31(1,30)-30(1,29) E	&81.50103	    &-3.2158	&273.44229&\\
	&343148.3182	&31(2,30)-30(1,29) E	&11.85121	    &-4.0532	&273.44229&\\
	&343149.3027	&17(5,12)-16(4,13) A	&1.87319	    &-4.5991	&107.80885&\\
	&343435.26	    &28(4,24)-27(4,23) E	&71.57138	    &-3.22767	&257.07971&\\
	&343443.944	    &28(4,24)-27(4,23) A	&71.58443	    &-3.22755	&257.0807&\\
	&343731.783	    &27(7,20)-26(7,19) E	&67.17309	    &-3.23857	&258.47228&\\
	&344051.371	    &28(19, 9)-27(19, 8) A	&40.45991	    &-3.47305	&478.72469&\\
	&344759.096	    &28(15,13)-27(15,12) A	&53.37173	    &-3.35008	&388.78295&\\
	&344762.59	    &28(15,13)-27(15,12) E	&53.3701	    &-3.35008	&388.77996&\\
	&345067.795	    &28(14,14)-27(14,13) E	&56.11927	    &-3.32712	&369.64085&\\
	&345069.059	    &28(14,15)-27(14,14) A	&56.11927	    &-3.32711	&369.64307&\\
	&345091.465	    &28(14,15)-27(14,14) E	&56.12379	    &-3.32699	&369.63767&\\
	&345095.559	    &19(13,7)-19(12,8) E	&2.84671	    &-4.45697	&224.16512&\\
	&345465.3445	&16(13,4)-16(12,5) E	&1.72784	    &-4.59986	&192.32292&\\
	&345466.962	    &28(13,16)-27(13,15) A	&58.67428	    &-3.30627	&351.85611&\\
	&345486.602	    &28(13,16)-27(13,15) E	&58.68009	    &-3.30616	&351.8513&\\
	&345509.0211	&16(13,3)-16(12,4) A	&1.72749	    &-4.59979	&192.33868&\\
	&345509.0211	&16(13,3)-16(12,4) A	&1.72749	    &-4.59979	&192.33868&\\
	&345974.664	    &28(12,16)-27(12,15) E	&61.03512	    &-3.28723	&335.43319&\\
	&345985.381	    &28(12,17)-27(12,16) A	&61.04559	    &-3.28711	&335.43442&\\
	&346001.616	    &28(12,17)-27(12,16) E	&61.03889	    &-3.2871	&335.43002&\\
	&354607.7638	&33(0,33)-32(1,32) E	&8.97859	    &-4.15769	&293.11551&\\
	&354607.7643	&33(1,33)-32(1,32) E	&92.83422	    &-3.14319	&293.11551&\\
	&354607.7648	&33(0,33)-32(0,32) E	&92.83422	    &-3.14319	&293.11551&\\
	&354607.7652	&33(1,33)-32(0,32) E	&8.97859	    &-4.15769	&293.11551&\\
	&354608.0913	&33(0,33)-32(1,32) A	&14.15735	    &-3.95991	&293.09811&\\
	&354608.0919	&33(1,33)-32(1,32) A	&87.67628	    &-3.16801	&293.09811&\\
	&354608.0923	&33(0,33)-32(0,32) A	&87.67628	    &-3.16801	&293.09811&\\
	&354608.0928	&33(1,33)-32(0,32) A	&14.15735	    &-3.95991	&293.09811&\\
	\hline  
	CH$_3$OCH$_3$ v=0   &342607.898	  &19(0,19)-18(1,18) AE	&110.05055	 &-3.48121	  &167.14232 &\\
	&342607.898	  &19(0,19)-18(1,18) AE	&165.07914	 &-3.4812	  &167.14232&\\
	&342607.971	  &19(0,19)-18(1,18) EE	&440.24256	 &-3.48117	  &167.14233&\\
	&342608.044	  &19(0,19)-18(1,18) AA	&275.16414	 &-3.48115	  &167.14233&\\
	&343753.32	  &17(2,16)-16(1,15) EA	&29.00781	 &-3.70791	  &143.69875&\\
	&343753.32	  &17(2,16)-16(1,15) EA	&58.02497	 &-3.70784	  &143.69875&\\
	&343754.216	  &17(2,16)-16(1,15) EE	&232.06664	 &-3.7079	  &143.69879&\\
	&343755.112	  &17(2,16)-16(1,15) AA	&87.01828	 &-3.70793	  &143.69883&\\
	&344357.816	  &19(1,19)-18(0,18) EA	&55.06275	 &-3.47427	  &167.17839&\\
	&344357.816	  &19(1,19)-18(0,18) EA	&110.1179	 &-3.4743	  &167.17839&\\
	&344357.929	  &19(1,19)-18(0,18) EE	&440.5119	 &-3.47426	  &167.1784&\\
	&344358.041	  &19(1,19)-18(0,18) AA	&165.17917	 &-3.4743	  &167.17697&\\
	&344512.176	  &11(3,9)-10(2,8) EA	&25.1312	 &-3.88603	  &72.78387&\\
	&344512.219	  &11(3,9)-10(2,8) AE	&12.56646	 &-3.886	  &72.78387&\\
	&344515.385	  &11(3,9)-10(2,8) EE	&100.53216	 &-3.88598	  &72.78388&\\
	&344518.572	  &11(3,9)-10(2,8) AA	&37.70482	 &-3.88591	  &72.78389&\\
	&356723.697	  &8(4,4)-7(3,5)  EE	&75.66091	 &-3.83277	  &55.26669&\\
	&356724.457	  &8(4,4)-7(3,5)  AA	&40.44545	 &-3.6788	  &55.26673&\\
	&356724.864	  &8(4,5)-7(3,5)  EA	&15.06151	 &-3.93171	  &55.2666&\\ 
	\hline
	CH$_3$COCH$_3$ v=0  
	&342485.2336	&17(17,0)-16(16,0) EE	&2209.9853	&-2.73394	&147.06155&\\
	&342485.6638	&40(1,39)-40(1,40) EA	&41.07562	&-4.2271	&411.34942&\\
	&342485.6638	&40(2,39)-40(0,40) EA	&41.07562	&-4.2271	&411.34942&\\
	&342486.2877	&40(1,39)-40(0,40) AE	&20.53915	&-4.22707	&411.3493&\\
	&342486.2877	&40(2,39)-40(1,40) AE	&61.61442	&-4.22709	&411.3493&\\
	&342594.8875	&17(17,1)-16(16,1) EE	&2210.2712	&-2.73347	&146.9497&\\
	&342596.9064	&18(14,4)-17(13,5) EA	&3.20786	&-4.99376	&147.13957&\\
	&342780.0345	&17(17,1)-16(16,0) AA	&828.91802	&-2.73273	&147.04764&\\
	&342780.0361	&17(17,0)-16(16,1) AA	&1381.37496	&-2.73278	&147.04764&\\
	&345037.4998	&18(15,3)-17(14,3) EE	&1589.86932	&-2.89143	&150.868&\\
	&345639.6192	&33(3,31)-32(3,30) EE	&3529.9523	&-2.76999	&298.03287&\\
	&345639.6192	&33(3,31)-32(2,30) EE	&592.06425	&-3.54539	&298.03287&\\
	&345639.6192	&33(2,31)-32(2,30) EE	&3529.9523	&-2.76999	&298.03287&\\
	&345639.6192	&33(2,31)-32(3,30) EE	&592.06425	&-3.54539	&298.03287&\\
	&345671.5235	&33(2,31)-32(2,30) AA	&1545.65029	&-2.73319	&297.98016&\\
	&345671.5236	&33(3,31)-32(3,30) AA	&2575.79467	&-2.73324	&297.98016&\\
	&345673.9538	&15(7,8)-14(6,9) EE	&115.73717	&-3.95008	&92.93574&\\
	&356902.7724	&18(17,1)-17(16,1) EE	&2100.85617	&-2.72634	&159.58433&\\
	&356905.4132	&15(6,9)-14(5,10) AA	&52.17466	&-4.05031	&89.51887&\\
	\hline
	H$_2$CS v=0  
	&342946.4239	&10(0,10)- 9(0,9)	    &27.19603	&-3.2161	&90.59115&\\
	&343203.2392	&10(5,6)- 9(5,5)	    &61.19507	&-3.34004	&419.17248&\\
	&343203.2392	&10(5,6)- 9(5,5)	    &61.19507	&-3.34004	&419.17248&\\
	&343309.8296	&10(4,7)- 9(4,6)	    &22.84414	&-3.29045	&301.07181&\\
	&343322.0819	&10(2,9)- 9(2,8)	    &26.10686	&-3.23243	&143.30653&\\
	&343409.9625	&10(3,8)- 9(3,7)	    &74.24497	&-3.2553	&209.09441&\\
	&343414.1463	&10(3,7)- 9(3,6)	    &74.24322	&-3.2553	&209.09476&\\
	&343813.1683	&10(2,8)- 9(2,7)	    &26.10949	&-3.23052	&143.37729&\\  
	\hline
	CH$_3$OH v$_t$=0 
	&342729.796	&13(1,12)-13(0,13) A	        &97.51607	    &-3.37356	&227.47312&\\
	&344109.039	&18(-2,17)-17(-3,15) E	        &21.23636	    &-4.16716	&419.39868&\\
	&344443.433	&19(1,19)-18(2,16) A	        &23.97181	    &-4.13614	&451.22691&\\
	&345903.916	&16(1,15)-15(2,14) A	        &28.51774	    &-3.98266	&332.6488&\\
	&345919.26	&18(3,15)-17(4,14) E	        &22.4084	    &-4.13699	&459.43023&\\
	\hline    
	CH$_3$OH v$_t$=1
	&344312.267	&10(2,9)-11(3,9) E	        &31.24862	    &-3.75266	&491.90986&\\
	&344312.374	&10(-2,8)- 11(-3,8) 	        &7.81628	    &-3.75242	&491.90961&\\
	\hline    
	H$_2$$^{13}$CO v=0 &343325.713	&5(1,5)- 4(1,4)	                    &78.29712	    &-2.95172	&61.27905&\\
	\hline
	H$_2$CCO v=0   &343172.5723	&17(0,17)-16(0,16)	  &34.27746	   &-3.33659	&148.30454&\\
	&343250.411	    &17(4,14)-16(4,13)	  &32.38127	   &-3.36101	&356.83923&\\
	&343376.133	    &17(2,16)-16(2,15)	  &33.8061	   &-3.34183	&200.53312&\\
	&343384.676	    &17(3,15)-16(3,14)	  &99.63928	   &-3.34948	&265.71292&\\
	&343387.579	    &17(3,14)-16(3,13)	  &99.63771	   &-3.34948	&265.71335&\\
	&343693.935	    &17(2,15)-16(2,14)	  &33.80534	   &-3.34063	&200.60564&\\  
	\hline
	$^{33}$SO v=0&343086.1019	&9(8)- 8(7), F=15/2-13/2	     &17.39217	&-3.29162	&78.02935&\\
	&343087.2979	&9(8)- 8(7), F=17/2-15/2	     &19.48739	&-3.29337	&78.03127&\\
	&343088.078	    &9(8)- 8(7), F=19/2-17/2	     &21.84011	&-3.28962	&78.03347&\\
	&343088.2949	&9(8)- 8(7), F=21/2-19/2	     &24.45445	&-3.28191	&78.03607&\\
	\hline
	OC$^{33}$S v=0 &343983.2336	&29-28	      &14.84195	&-3.92382	&247.66956&\\
	\hline
	t-HCOOH v=0 
	&343952.4127	&15(1,14)-14(1,13)	 &29.03379	    &-3.35303	&136.28091&\\
	&345030.5959	&16(0,16)-15(0,15)	 &31.05632	    &-3.34686	&143.05228&\\
	\hline
	SO$_2$  v=0  
	&345338.5377	&13(2,12)-12(1,11)	    &13.40994	&-3.62327	&92.98367&\\
	&356755.1899	&10(4,6)-10(3,7)	    &13.03515	&-3.48406	&89.83365&\\
	\hline
	HNCCC v=0 
	&345343.0169	&37-36, F=38-37	   &1219.00467	&-2.1198	&314.95193&\\
	&345343.017	    &37-36, F=37-36	   &1186.6028	&-2.12007	&314.95193&\\
	&345343.0172	&37-36, F=36-35	   &1154.79626	&-2.12013	&314.95193&\\
	\hline
	H$^{13}$CN v=0  &345338.1613	&4 - 3, F = 4 - 4	   &2.22765	    &-3.92573	&41.43538&\\
	&345339.6586	&4 - 3, F = 3 - 2	   &25.45912	&-2.75858	&41.43545&\\
	&345339.7693	&4 - 3, F = 4 - 3	   &33.41513	&-2.74963	&41.43546&\\
	&345339.8148	&4 - 3, F = 5 - 4	   &43.56554	&-2.72158	&41.43546&\\
	&345341.7462	&4 - 3, F = 3 - 3	   &2.22761	    &-3.81658	&41.43555&\\
	\hline
	C$_{2}$H$_{5}$OH v=0  
	&345229.2848	&21(1,21) - 20(1,20), g+	        &33.33569	    &-3.4303	&241.55353&\\
	&345408.1651	&21(0,21) - 20(0,20), g-	        &32.93761	    &-3.43484	&246.20853&\\
	&354363.2335	&20(3,17) - 19(2,17), g+	        &38.6078	    &-3.31183	&244.98529&\\
	\hline
	NS v=0  &345823.288	    &15/2-13/2, $\Omega$=1/2, F=17/2-15/2,l=e	    &27.33931	&-3.13599	&69.68377&\\
	&345823.288	    &15/2-13/2,$ \Omega$=1/2, F=17/2-15/2,l=e	    &23.8666	&-3.14383	&69.68579&\\
	&345824.13	    &15/2-13/2,$ \Omega$=1/2, F=13/2-11/2,l=e	    &20.83008	&-3.14494	&69.68755&\\
	\hline
	HC$_3$N v=0  &345609.01	     &38-37	    &529.12907	&-2.48124	&323.49156&\\
	\hline
	SO v=0  &344310.612	    &8(8)- 7(7)	&18.55549	   &-3.28519	&87.48155&\\
	\hline
	$^{34}$SO$_2$ v=0  
	&344581.0445	&19(1,19)-18(0,18)	   &42.23967	&-3.28754	&167.65667&\\
	&344807.9147	&13(4,10)-13(3,11)	   &17.92316	&-3.49928	&121.63281&\\
	&344987.5847	&15(4,12)-15(3,13)	   &21.20438	&-3.48559	&148.34497&\\
	&344998.1602	&11(4,8)-11(3,9)	       &14.6664	    &-3.51602	&98.62644&\\
	&345168.6641	&8(4,4)-8(3,5)	       &9.76929	    &-3.56056	&71.047&\\
	&345285.6199	&9(4,6)-9(3,7)	       &11.41398	&-3.54085	&79.31737&\\
	&345519.6563	&7(4,4)-7(3,5)	       &8.08591	    &-3.58701	&63.69764&\\
	&345553.0927	&6(4,2)-6(3,3)	       &6.34216	    &-3.63023	&57.26899&\\
	&345651.2934	&5(4,2)-5(3,3)	       &4.49068	    &-3.70723	&51.76028&\\
	&345678.7871	&4(4,0)-4(3,1)	       &2.44173	    &-3.88459	&47.1706&\\
	&345929.349	    &17(4,14)-17(3,15)	   &24.49457	&-3.4721	&178.77039&\\
	\hline
	$^{13}$CH$_3$OH v$_t$=0  &344040.629	&8(-3,6)- 9(-2,8)	&2.17479	&-4.21727	&144.4915&\\
	&345132.599	 &4 (0,4)- 3 (-1,3)	    &1.54925	 &-4.08423	&35.76015&\\
	&354445.952	 &4 (1,3)- 3 (0,3)	    &2.21298	 &-3.89468	&43.71183&\\
	&356873.814	 &13(2,12)- 12(3,9)    &4.04181	 &-4.10131	&243.83835&\\
	\hline
	NH$_2$CHO v=0 &343083.117	    &16(3,13) - 15(3,12)	&201.8469	&-2.54136	&165.99525&\\
	&343196.9863	&17(1,17) - 16(1,16)	&221.24033	&-2.52664	&151.99629&\\
	&345181.258	    &17(0,17) - 16(0,16)	&221.41017	&-2.51879	&151.58895&\\
	&345325.3906	&16(1,15) - 15(1,14)	&207.83422	&-2.52018	&145.15586&\\
	&356713.7586	&17(2,16) - 16(2,15)	&218.95428	&-2.48082	&166.78985&\\
	\hline
	OS$^{18}$O v=0 &344873.823	    &14(4,10)-14(3,11)  &18.98449	 &-3.50508	&136.08524&\\
	\hline
	$^{33}$SO$_2$ v=0 
	&345584.6906	&19(1,19)-18(0,18), F=35/2-33/2	&38.45552	&-3.28975	&170.23471&\\
	&345584.7147	&19(1,19)-18(0,18), F=41/2-39/2	    &45.06702	&-3.28779	&170.23486&\\
	&345585.2427	&19(1,19)-18(0,18), F=37/2-35/2	    &40.54693	&-3.29023	&170.23518&\\
	&345585.2568	&19(1,19)-18(0,18), F=39/2-37/2	    &42.7423	&-3.2896	&170.23518&\\
	\hline
\end{longtable}
During the preparation of this work the authors used Deepseek in order to improve language and readability of the manuscript. 
After using this tool, the authors reviewed and edited the content as needed and take full responsibility for the content of the publication.

\bibliography{ms2025-0541}{}
\bibliographystyle{raa}
\end{document}